%% file: main.tex
\documentclass{LMCS}

\usepackage{amssymb}
\usepackage{amsmath}
\usepackage{amscd}
\usepackage{latexsym}
\usepackage{epic}
\usepackage{eepic}
\usepackage{psfrag}
\usepackage{psfig}
\usepackage{graphicx}
\usepackage{algorithm}
\usepackage{algorithmic}
\usepackage{example}
\usepackage{luca}
\usepackage{macro}
\usepackage{subfigure}
\usepackage[usenames]{color}
\usepackage{color}
\usepackage{colortbl}
\usepackage[all]{xy}
\usepackage{enumerate,hyperref}

\def\doi{6 (3:13) 2010}
\lmcsheading%
{\doi}
{1--27}
{}
{}
{Oct.~20, 2009}
{Sep.~\phantom01, 2010}
{}

\begin{document}

\title[Algorithms for Game Metrics]
{Algorithms for Game Metrics\rsuper*}

\author[K.~Chatterjee]{Krishnendu Chatterjee\rsuper a}	
\address{{\lsuper a}IST Austria (Institute of Science and Technology Austria)}
\email{krish.chat@ist.ac.at}

\author[L.~de Alfaro]{Luca de Alfaro\rsuper b}	
\address{{\lsuper{b,d}}Computer Science Department, University of California, Santa Cruz}
\email{\{luca,vishwa\}@soe.ucsc.edu}

\author[R.~Majumdar]{Rupak Majumdar\rsuper c}	
\address{{\lsuper c}Department of Computer Science, University of California, Los Angeles}
\email{rupak@cs.ucla.edu}

\author[V.~Raman]{Vishwanath Raman\rsuper d}	





\titlecomment{{\lsuper*}A preliminary version of this paper titled 
``Algorithms for Game Metrics" appeared in the IARCS Annual Conference 
on Foundations of Software Technology and Theoretical Computer Science, 
December 2008. This is a version with full proofs and extensions.}



\begin{abstract}
\noindent Simulation and bisimulation metrics for stochastic systems 
provide a quantitative generalization of the classical simulation and
bisimulation relations. 
These metrics capture the similarity of states with respect to
quantitative specifications written in the quantitative $\mu$-calculus
and related probabilistic logics.
We first show that the metrics provide a bound for the 
difference in {\em long-run average\/} and {\em discounted average\/}
behavior across states, indicating that the metrics can be used both
in system verification, and in performance evaluation. 
For turn-based games and MDPs, we provide a polynomial-time algorithm
for the computation of the one-step metric distance between states. 
The algorithm is based on linear programming; it improves on the
previous known exponential-time algorithm based on a reduction to the 
theory of reals.
We then present PSPACE algorithms for both the decision problem and the
problem of approximating the metric distance between two states,
matching the best known algorithms for Markov chains. 
For the bisimulation kernel of the metric our algorithm works in time 
$\calo(n^4)$ for both turn-based games and MDPs; improving the 
previously best known $\calo(n^9\cdot\log(n))$ time algorithm for MDPs. 

For a concurrent game $G$, we show that computing the exact distance
between states is at least as hard as computing the value of
concurrent reachability games and
the square-root-sum problem in computational geometry.
We show that checking whether the metric distance is bounded by a 
rational $r$, can be done via a reduction to the theory of 
real closed fields, involving a formula with three quantifier 
alternations, yielding $\calo(|G|^{\calo(|G|^5)})$ time 
complexity, improving the previously known reduction, which yielded  
$\calo(|G|^{\calo(|G|^7)})$ time complexity.
These algorithms can be iterated to approximate the metrics using
binary search. 
\end{abstract}

\maketitle\vfill

\setcounter{page}{1}

\input{introduction}
\input{definitions}

\input{metrics}
\input{average}
\input{turnbased}
\input{concurrent}
\input{discussion}

\subsection*{Acknowledgments.}
The first, second and fourth author were supported in part by the 
National Science Foundation grants CNS-0720884 and CCR-0132780. 
The third author was supported in part by the National Science
Foundation grants CCF-0427202, CCF-0546170.
We would like to thank the reviewers for their detailed comments
that helped us make the paper better.

\bibliographystyle{plain}
\bibliography{dvlab}

\end{document}

%% file: introduction.tex
\section{Introduction}

System metrics constitute a quantitative generalization of system
relations. 
The bisimulation relation captures state {\em equivalence:\/} two
states $s$ and $t$ are bisimilar if and only if they cannot be
distinguished by any formula of the
$\mu$-calculus \cite{BCG88}.
The bisimulation {\em metric\/} captures the {\em degree of
difference\/} between two states: the bisimulation distance between
$s$ and $t$ is a real number that provides a tight bound for the
difference in value of formulas of the {\em quantitative\/}
$\mu$-calculus at $s$ and $t$ \cite{DGJP99}. 
A similar connection holds between the simulation relation and the
simulation metric. 

The classical system relations are a basic tool in the study of 
{\em boolean\/} properties of systems, that is, the properties that
yield a truth value. 
As an example, if a state $s$ of a transition system can reach a set
of target states $R$, written $s \sat \diam R$ in temporal logic, and
$t$ can simulate $s$, then we can conclude $t \sat \diam R$. 
System metrics play a similarly fundamental role in the study of the
quantitative behavior of systems. 
As an example, if a state $s$ of a Markov chain can reach a set of
target states $R$ with probability $0.8$, written 
$s \sat \P_{\geq 0.8} \diam R$, and if the metric simulation distance
from $t$ to $s$ is 0.3, then we can conclude 
$t \sat \P_{\geq 0.5} \diam R$. 
The simulation relation is at the basis of the notions of system
refinement and implementation, where qualitative properties are
concerned. 
In analogous fashion, simulation metrics provide a notion of
approximate refinement and implementation for quantitative properties. 

We consider three classes of systems: 
\begin{itemize}
\item {\em Markov decision processes.}  In these systems there is one
  player.  At each state, the player can choose a move; the current
  state and the move determine a probability distribution over
  the successor states.
\item {\em Turn-based games.}  In these systems there are two
  players.  At each state, only one of the two players can choose a move; 
  the current state and the move determine a probability distribution over 
  the successor states.  
\item{\em Concurrent games.}  In these systems there are two
  players.  At each state, both players choose moves simultaneously
  and independently; the current state and the chosen moves determine  a
  probability distribution over the successor states.
\end{itemize}
System metrics were first studied for Markov chains and Markov decision
processes (MDPs)
\cite{DGJP99,vanBreugelCONCUR01,vanBreugel-icalp2001,DGJP02,RadhaLICS02},
and they have recently been extended to two-player turn-based and
concurrent games \cite{dAMRS07}. 
The fundamental property of the metrics is that they provide a tight
bound for the difference in value that formulas belonging to
quantitative specification languages assume at the states of a
system. 
More precisely, let $\qmu$ indicate the 
{\em quantitative $\mu$-calculus,} a specification language in which
many of the classical specification properties, including reachability
and safety properties, can be written \cite{dAM04}. 
The metric bisimulation distance between two states $s$ and $t$,
denoted $[s \priobis_g t]$, has the property that 
$[s \priobis_g t] = \sup_{\varphi \in \qmu} |\varphi(s) - \varphi(t)|$, where
$\varphi(s)$ and $\varphi(t)$ are the values $\varphi$ assumes at $s$ and $t$. 
To each metric is associated a {\em kernel:\/} the kernel of a metric
$d$ is the relation that relates the pairs of states that have
distance~0; to each metric corresponds a metric kernel relation. 
The kernel of the simulation metric is {\em probabilistic simulation;\/}
the kernel of the bisimulation metric is 
{\em probabilistic bisimulation\/} \cite{SL94}. 


\smallskip\noindent{\bf Metric as bound for discounted and long-run 
average payoff.} 
Our first result is that the metrics developed in \cite{dAMRS07}
provide a bound for the difference in {\em long-run average\/}
and {\em discounted average\/} properties across states of a system. 
These average rewards play a central role in the theory of stochastic
games, and in its applications to optimal control and economics
\cite{Bertsekas95,FilarVrieze97}.
Thus, the metrics of \cite{dAMRS07} are useful both for system
verification, and for performance evaluation, supporting our belief
that they constitute  the canonical metrics for the study of the
similarity of states in a game.
We point out that it is possible to define a discounted version 
$\metr{\priobis_g}^\dfactor$ of the game bisimulation
metric; however, we show that this discounted metric does 
{\em not\/} provide a bound for the difference in discounted values.

\smallskip\noindent{\bf Algorithmic results.}
Next, we investigate algorithms for the computation of the metrics.
The metrics can be computed in iterative fashion, following the
inductive way in which they are defined. 
A metric $d$ can be computed as the limit of a monotonically
increasing sequence of approximations $d_0$, $d_1$, $d_2$, \ldots, 
where $d_0(s,t)$ is the difference in value that variables can have at
states $s$ and $t$.
For $k \geq 0$, $d_{k+1}$ is obtained from $d_k$ via 
$d_{k+1} = H(d_k)$, where the operator $H$ depends on the metric
(bisimulation, or simulation), and on the type of system. 
Our main results are as follows:
\begin{enumerate}
\item {\em Metrics for turn-based games and MDPs.}
We show that for turn-based games, and MDPs, the one-step metric
operator $H$ for both bisimulation and simulation can be computed in
polynomial time, via a reduction to linear programming (LP). 
The only previously known algorithm, which can be inferred from
\cite{dAMRS07}, had EXPTIME complexity and relied on a reduction to
the theory of real closed fields;
the algorithm thus had more a complexity-theoretic, than a practical, value.
The key step in obtaining our polynomial-time algorithm consists in
transforming the original $\sup$-$\inf$ \emph{non-linear} optimization problem
(which required the theory of reals) into a quadratic-size
$\inf$ \emph{linear} optimization problem that can be solved via LP. 
We then present PSPACE algorithms for both the decision problem of the 
metric distance between two states and for the problem of computing
the approximate metric distance between two states for turn-based 
games and MDPs. 
Our algorithms match the complexity of the best known algorithms for 
the sub-class of Markov chains \cite{vBSW08}.

\item {\em Metrics for concurrent games.}
For concurrent games, our algorithms for the $H$ operator still rely on
decision procedures for the theory of real closed fields, leading to 
an EXPTIME procedure. 
However, the algorithms that could be inferred from \cite{dAMRS07} had 
time-complexity $\calo(|G|^{\calo(|G|^7)})$, where $|G|$ is the size of a
game;
we improve this result by presenting algorithms with 
$\calo(|G|^{\calo(|G|^5)})$
time-complexity. 

\item {\em Hardness of metric computation in concurrent games.}
We show that computing the exact distance of states of
concurrent games is at least as hard as computing the value of
concurrent reachability games \cite{EY06,crg-tcs07}, which is known to be
at least as hard as solving the square-root-sum problem in
computational geometry \cite{GareyGrahamJohnson76}. 
These two problems are known to lie in PSPACE, and have resisted
many attempts to show that they are in NP. 

\item {\em Kernel of the metrics.} We present polynomial time 
algorithms to compute the simulation and bisimulation kernel of the
metrics for turn-based games and MDPs.
Our algorithm for the bisimulation kernel
of the metric runs in time $\calo(n^4)$ 
(assuming a constant number of moves) as compared to the previous known 
$\calo(n^9\cdot\log(n))$ algorithm of \cite{ZhangH07} for MDPs, where $n$ is
the size of the state space.
For concurrent games the simulation and the bisimulation kernel can be 
computed in time $\calo(|G|^{\calo(|G|^3)})$, where $|G|$ is the size of a
game.

\end{enumerate}

Our formulation of probabilistic simulation and bisimulation differs 
from the one previously considered for MDPs in \cite{Baier96}: there, the 
names of moves (called ``labels'') must be preserved by simulation and 
bisimulation, so that a move from a state has at most one candidate 
simulator move at another state. 
Our problem for MDPs is closer to the one considered in
\cite{ZhangH07}, where labels must be preserved, but where a label can
be associated with multiple probability distributions (moves).

For turn-based games and MDPs, the algorithms for probabilistic
simulation and bisimulation can be obtained from the LP
algorithms that yield the metrics. 
For probabilistic simulation, the algorithm we obtain coincides with
the algorithm previously published in \cite{ZhangH07}. 
The algorithm requires the solution of feasibility-LP problems with a
number of variables and inequalities that is quadratic in the size of
the system. 
For probabilistic bisimulation, we are able to improve on
this result by providing an algorithm that requires the solution of
feasibility-LP problems that have linearly many variables and
constraints. 
Precisely, as for ordinary bisimulation, the kernel is computed via iterative
refinement of a partition of the state space \cite{Milner90}. 
Given two states that belong to the same partition, to decide whether 
the states need to be split in the next partition-refinement step, we
present an algorithm that requires the solution of a feasibility-LP
problem with a number of variables equal to the number of moves
available at the states, and number of constraints linear in the
number of equivalence classes. 
Overall, our algorithm for bisimulation runs in time $\calo(n^4)$ 
(assuming a constant number of moves),
considerably improving the $\calo(n^9\cdot\log(n))$ algorithm of 
\cite{ZhangH07} for MDPs, and providing
for the first time a polynomial algorithm for turn-based games.



%% file: definitions.tex
\section{Definitions}

\noindent{\bf Valuations.}
Let $[\lb, \rb]\subs \reals$ be a fixed, non-singleton real interval.
Given a set of states $S$, a {\em valuation over $S$\/} 
is a function $\valu: S \mapsto [\lb, \rb]$
associating with every state $s \in S$ a value $\lb \leq \valu(s) \leq \rb$;
we let $\valus$ be the set of all valuations. 
For $c \in [\lb,\rb]$, we denote by $\imeanbb{c}$ the constant valuation
such that $\imeanbb{c}(s) = c$ at all $s \in S$. 
We order valuations pointwise: for $\valu, \valub \in \valus$, we
write $\valu \leq \valub$ iff $\valu(s) \leq \valub(s)$ at all $s \in S$;
we remark that $\valus$, under $\leq$, forms a lattice. 
Given $a,b \in \reals$, we write $a \imax b = \max\set{a,b}$, and
$a \imin b = \min\set{a,b}$; we also let
$a \oplus b = \min \set{1, \max \set{0, a + b}}$ and
$a \ominus b = \max \set{0, \min \set{1, a - b}}$.
We extend $\imin, \imax, +, -, \oplus, \ominus$ to valuations by 
interpreting them in pointwise fashion.

\bigskip \noindent{\bf Game structures.}
For a finite set $A$, let $\distr(A)$ denote the set of probability
distributions over $A$. 
We say that $p \in \distr(A)$ is {\em deterministic\/} if there is 
$a \in A$ such that $p(a) = 1$. 
We assume a fixed finite set $\vars$ of {\em observation variables\/}.

A  (two-player, concurrent) {\em game structure\/} 
$\game=\tuple{S,\int{\cdot},\moves,\mov_1,\mov_2,\trans}$
consists of the  following components \cite{ATL02,crg-tcs07}:
\begin{itemize}

\item A finite set $S$ of states.

\item A variable interpretation $\int{\cdot}: \vars \mapsto 
  [\lb,\rb]^S$, which associates with each variable $\varx \in
  \vars$ a valuation $\int{v}$. 

\item A finite set $\moves$ of moves.

\item Two move assignments 
  $\mov_1,\mov_2$: $S\mapsto 2^\moves \setm \set{\emptyset}$.  
  For $i\in\{1,2\}$, the assignment $\mov_i$ associates with each
  state $s \in S$ the nonempty set $\mov_i(s)\subseteq\moves$ of moves
  available to player $i$ at state~$s$.

\item A probabilistic transition function $\trans$: 
  $S\times\moves\times\moves\mapsto\distr(S)$, that gives the probability 
  $\trans(s,a_1,a_2)(t)$ of a transition from $s$ to $t$ when player 1
  plays move $a_1$ and player 2 plays move~$a_2$. 

\end{itemize}
At every state $s\in S$, player~1 chooses a move $a_1\in\mov_1(s)$, 
and simultaneously and independently player~2 chooses a move 
$a_2\in\mov_2(s)$.  
The game then proceeds to the successor state $t\in S$ with probability
$\trans(s,a_1,a_2)(t)$.
We let $\dest(s,a_1,a_2) = \set{t \in S \mid \trans(s,a_1,a_2)(t) > 0}$.
The {\em propositional distance} $\propdist(s,t)$ between two states 
$s,t\in S$ is the maximum difference in the valuation of any variable:
\[
\propdist(s,t) = \max_{\varx\in\vars} \vert\int{\varx}(s) - \int{\varx}(t)\vert 
\eqpun .
\]
The kernel of the propositional distance induces an equivalence on
states: for states $s,t$, we let $s\loceq t$ if $\propdist(s,t) = 0$.
In the following, unless otherwise noted, the definitions refer to a game
structure with components 
$\game=\tuple{S,\int{\cdot},\moves,\mov_1,\mov_2,\trans}$. 
We indicate the opponent of a player $\ii\in\set{1,2}$ by 
$\jj = 3 - \ii$.
We consider the following subclasses of game structures. 

\smallskip
\noindent{\bf Turn-based game structures.} 
A game structure $\game$ is {\em turn-based\/} if we can
write $S = S_1 \union S_2$ with $S_1 \inters S_2 = \emptyset$ 
where $s \in S_1$ implies $|\mov_2(s)| = 1$, and $s \in S_2$ implies
$|\mov_1(s)| = 1$, and further, there exists a special variable
$\turn \in \vars$, such that $\int{\turn}{s} = \theta_1$ iff $s \in S_1$, and 
$\int{\turn}{s} = \theta_2$ iff $s \in S_2$. 

\smallskip
\noindent{\bf Markov decision processes.}
For $i \in \set{1,2}$, we say that a structure is an $i$-MDP
if $\forall s \in S$, $\vert \mov_{\jj}(s) \vert = 1$. 
For MDPs, we omit the (single) move of the player
without a choice of moves, and 
write $\delta(s,a)$ for the transition function. 

\bigskip \noindent{\bf Moves and strategies.}
A {\em mixed move\/} is a probability distribution over the moves
available to a player at a state. 
We denote by $\dis_i(s) \subs \distr(\moves)$ the set of mixed moves 
available to player~$i \in \set{1,2}$ at $s \in S$, where: 
\[
  \dis_i(s) = \set{\dis \in \distr(\moves) \mid 
  \dis(a) > 0 \text{\textit{ implies }} a \in \mov_i(s)} \eqpun .
\]
The moves in $\moves$ are called {\em pure moves.\/} 
We extend the transition function to mixed moves by defining, for 
$s \in S$ and $x_1 \in \dis_1(s)$, $x_2 \in \dis_2(s)$,
\[
  \trans(s,x_1,x_2)(t) =
  \sum_{a_1 \in \mov_1(s)} \: \sum_{a_2 \in \mov_2(s)} \:
  \trans(s,a_1,a_2)(t) \cdot x_1(a_1) \cdot x_2(a_2) \eqpun . 
\]
A {\em path\/} $\path$ of $\game$ is an infinite sequence 
$s_0, s_1, s_2,...$ of states in $s \in S$, such that for all $k \ge 0$, 
there exist moves $a_1^k \in \mov_1(s_k)$ and $a_2^k \in \mov_2(s_k)$
with $\trans(s_k, a_1^k, a_2^k)(s_{k+1}) > 0$.
We write $\Paths$ for the set of all paths, and $\Paths_s$ for the set 
of all paths starting from state $s$.

A \emph{strategy} for player $i \in \{1, 2\}$ is a function $\stra_i:
S^+ \mapsto \distr(\moves)$ that associates with every non-empty
finite sequence $\path \in S^+$ of states, representing the history of
the game, a probability distribution $\stra_i(\path)$, which is used
to select the next move of player~$i$;
we require that for all $\path \in S^*$ and
states $s \in S$, if $\stra_i(\path s)(a) > 0$, then $a \in \mov_i(s)$.
We write $\Stra_i$ for the set of strategies for player~$i$.
Once the starting state $s$ and the strategies $\straa$ and $\strab$
for the two players have been chosen, the game is reduced to an
ordinary stochastic process, denoted $\game_s^{\straa, \strab}$, which
defines a probability distribution on the set $\Paths$ of paths.
We denote by $\Pr_s^{\straa,\strab}(\cdot)$ the probability of a measurable
event (sets of paths) with respect to this process, and denote by
$\E_s^{\straa,\strab}(\cdot)$ the associated expectation operator.
For $k \geq 0$, we let $X_k: \Paths \to S$ be 
the random variable denoting the $k$-th state along a path.

\smallskip\noindent{\bf One-step expectations and predecessor operators.}
Given a valuation $\valu \in \valus$, a state $s \in S$, 
and two mixed moves $x_1 \in \dis_1(s)$ and $x_2 \in \dis_2(s)$, we define
the expectation of $\valu$ from $s$ under $x_1, x_2$ by,
\[
  \E^{x_1,x_2}_s (\valu) = 
  \sum_{t \in S} \, \trans(s,x_1,x_2)(t) \, \valu(t) \eqpun .
\]
For a game structure $\game$, for $i \in \set{1,2}$ we define the 
{\em valuation transformer\/} $\pre_i: \valus \mapsto \valus$ 
by, for all $\valu \in \valus$ and $s \in S$ as,
\[
  \pre_i(\valu)(s) =
  \Sup_{x_\ii \in \dis_\ii(s)} \; \Inf_{x_\jj \in \dis_\jj(s)} 
  \E^{x_\ii,x_\jj}_s (\valu) \eqpun . 
\]
Intuitively, $\pre_i(\valu)(s)$ is the maximal expectation player $i$
can achieve of $\valu$ after one step from $s$: this is the standard
``one-day'' or ``next-stage'' operator of the theory of repeated games
\cite{FilarVrieze97}. 

\subsection{Quantitative $\mu$-calculus}

We consider the set of properties expressed by the {\em
quantitative $\mu$-calculus\/} ($\qmu$). 
As discussed in 
\cite{Kozen83,dAM04,IverMorgan}, a large set of properties
can be encoded 
in $\qmu$, spanning from basic properties such as maximal reachability
and safety probability, to the maximal probability of satisfying a
general $\omega$-regular specification. 

\smallskip
\noindent{\bf Syntax.}
The syntax of quantitative $\mu$-calculus is defined with respect to
the set of observation variables $\vars$ as well as 
a set $\mvars$ of {\em calculus variables,} which are
distinct from the observation variables in $\vars$. 
The syntax is given as follows: 
\begin{align*}
  \varphi\ ::=\ &
    c \mid 
    \varx \mid 
    \mvar \mid
    \neg \varphi \mid 
    \varphi\vee\varphi \mid 
    \varphi \wedge\varphi \mid  
    \varphi \oplus c \mid
    \varphi \ominus c \mid
    \qpre_1(\varphi) \mid 
    \qpre_2(\varphi) \mid
    \mu \mvar.\, \varphi \mid
    \nu \mvar.\, \varphi
\end{align*}
for constants $c \in [\lb,\rb]$, observation variables $\varx \in \vars$, 
and calculus variables $\mvar \in \mvars$. 
In the formulas $\mu \mvar.\, \varphi$ and $\nu \mvar.\, \varphi$, we
furthermore require that all occurrences of the bound variable $\mvar$ in 
$\varphi$ occur in the scope of an even number of occurrences of the 
complement operator $\neg$. 
A formula $\varphi$ is {\em closed\/} if every calculus variable $\mvar$
in $\varphi$ occurs in the scope of a quantifier $\mu \mvar$ or $\nu
\mvar$.
From now on, with abuse of notation, we denote by $\qmu$ the set of closed
formulas of $\qmu$.  
A formula is a {\em player~$i$ formula,} for $i \in \set{1,2}$, 
if $\varphi$ does not contain the $\qpre_{\jj}$ operator; we denote with
$\qmu_i$ the syntactic subset of $\qmu$ consisting only of closed player~$i$
formulas. 
A formula is in {\em positive form\/} if the negation appears only in
front of constants and observation variables, 
i.e., in the context $\neg c$ and $\neg \varx$; we 
denote with $\qmu^{+}$ and $\qmu_i^{+}$ the subsets of $\qmu$ and $\qmu_i$
consisting only of positive formulas.  

We remark that the fixpoint operators $\mu$ and $\nu$ will not be
needed to achieve our results on the logical characterization of game
relations. 
They have been included in the calculus because they allow the
expression of many interesting properties, such as safety,
reachability, and in general, $\omega$-regular properties. 
The operators $\oplus$ and $\ominus$, on the other hand, are 
necessary for our results. 

\smallskip\noindent
{\bf Semantics.}
A variable valuation  $\xenv$: $\mvars\mapsto\valus$ is a function that maps 
every variable $\mvar\in\mvars$ to a valuation in~$\valus$.
We write $\xenv[\mvar\mapsto\valu]$ for the valuation that agrees
with $\xenv$ on all variables, except that $\mvar$ is mapped to~$\valu$. 
Given a game structure $\game$ and a variable valuation $\xenv$, every formula $\varphi$ of the
quantitative $\mu$-calculus defines a valuation 
$\sem{\varphi}^\game_{\xenv}\in\valus$ (the superscript $\game$ is
omitted if the game structure is clear from the context):
\begin{align*}
  & \sem{c}_{\xenv} = \imeanbb{c}  &
  & \sem{\varx}_{\xenv} = \int{\varx} \\
  & \sem{\mvar}_{\xenv} = \xenv(\mvar) &
  & \sem{\neg \varphi}_\xenv = \imeanbb{1} - \sem{\varphi}_\xenv \\
  & \textstyle \sem{\varphi {\oplus \brace \ominus} c}_\xenv = \textstyle \sem{\varphi}_\xenv {\oplus \brace \ominus} \imeanbb{c} &
  & \textstyle \sem{\varphi_1\,{\vee\brace\wedge}\,\varphi_2}_{\xenv} =
    \textstyle \sem{\varphi_1}_{\xenv} \,{\imax\brace\imin}\, 
    \sem{\varphi_2}_{\xenv} \\
  & \sem{\qpre_i(\varphi)}_\xenv = \pre_i(\sem{\varphi}_\xenv) &
  & \textstyle \sem{{\mu\brace\nu}\mvar.\, \varphi}_{\xenv} =
	\textstyle {\inf\brace\sup}\set{\valu \in \valus \mid 
    	\valu=\sem{\varphi}_{\xenv[\mvar\mapsto \valu]}}
\end{align*}
where $i \in \set{1,2}$. 
The existence of the fixpoints is guaranteed by the monotonicity 
and continuity of all operators and can be computed by Picard 
iteration \cite{dAM04}.
If $\varphi$ is closed, $\sem{\varphi}_\xenv$ is independent of $\xenv$, and
we write simply $\sem{\varphi}$. 

\smallskip\noindent{\bf Discounted quantitative $\mu$-calculus.}
A {\em discounted\/} version of the $\mu$-calculus was 
introduced in \cite{luca-icalp-disc-03}; we call this $\dmu$.
Let $\Lambda$ be a finite set of discount parameters
that take values in the interval $[0, 1)$.
The discounted $\mu$-calculus extends $\qmu$ by introducing discounted
versions of the player $\qpre$ modalities.
The syntax replaces $\qpre_i(\varphi)$ for player $i \in \set{1, 2}$ with 
its discounted variant, $\lambda \cdot \qpre_i(\varphi)$, where 
$\lambda \in \Lambda$ is
a discount factor that discounts one-step valuations.
Negation in the calculus is defined as $\neg (\lambda \cdot 
\qpre_1(\varphi)) = (1 - \lambda) + \lambda \cdot \qpre_2(\neg \varphi)$.
This leads to two additional pre-modalities for the players, 
$(1 - \lambda) + \lambda \cdot \qpre_i(\varphi)$.

%% file: metrics.tex
\paragraph{Game bisimulation and simulation metrics.}

A {\em directed metric\/} is a function $d: S^2 \mapsto \reals_{\geq 0}$ 
which satisfies $d(s,s) = 0$ and the \emph{triangle inequality} 
$d(s,t) \leq d(s,u) + d(u,t)$ for all $s,t,u \in S$. 
We denote by $\metrsp \subseteq S^2 \mapsto \reals$ the space of all 
directed metrics;
this space, ordered pointwise, forms a lattice which we indicate with
$(\metrsp, \leq)$.
Since $d(s, t)$ may be zero for $s \ne t$, these functions are {\em pseudo-metrics\/}
as per prevailing terminology~\cite{vanBreugelCONCUR01}.
In the following, we omit ``directed'' and simply say metric when the
context is clear.
For a metric $d$, we indicate with $C(d)$ the set of valuations
$k\in\valus$ where $k(s) - k(t) \leq d(s,t)$ for every $s,t\in S$.
A metric transformer $H_{\priosim_1}: \metrsp \mapsto \metrsp$ is
defined as follows, for all $d \in \metrsp$ and $s, t \in S$:
\begin{equation} \label{eq-game-met}
H_{\priosim_1}(d)(s,t) = \propdist(s, t) \imax 
  \Sup_{k \in C(d)} \bigl( 
  \pre_1(k)(s) - \pre_1(k)(t) \bigr) \eqpun . 
\end{equation}
The {\em player~1 game simulation metric\/} $[\priosim_1]$ is the 
least fixpoint of $H_{\priosim_1}$; the 
{\em game bisimulation metric\/} $[\priobis_1]$ is the least 
symmetrical fixpoint of $H_{\priosim_1}$ and is defined as follows,
for all $d \in \metrsp$ and $s, t \in S$:
\begin{equation} \label{eq-game-bis-met}
H_{\priobis_1}(d)(s,t) = H_{\priosim_1}(d)(s,t) \imax H_{\priosim_1}(d)(t,s)
\eqpun .
\end{equation}
The operator $H_{\priosim_1}$ is monotonic, non-decreasing and continuous 
in the lattice $(\metrsp, \leq)$.
We can therefore compute $H_{\priosim_1}$ using Picard iteration; we
denote by $\metr{\priosim_1^n} = H_{\priosim_1^n} (\imeanbb{0})$ 
the $n$-iterate of this.
%
From the determinacy of concurrent games with respect to
$\omega$-regular goals~\cite{Martin98}, we have that the game bisimulation metric 
is {\em reciprocal\/}, in that $[\priobis_1] = [\priobis_2]$;
we will thus simply write $[\priobis_g]$.
Similarly,  for all $s, t \in S$ we have $[s \priosim_1 t] = [t \priosim_2 s]$.

The main result in \cite{dAMRS07} about these metrics is that they 
are logically characterized by the quantitative $\mu$-calculus of
\cite{dAM04}.
We omit the formal definition of the syntax and semantics of the 
quantitative $\mu$-calculus; we refer the reader to \cite{dAM04} for 
details.
Given a game structure $\game$, every closed formula $\varphi$ of the 
quantitative $\mu$-calculus defines a valuation 
$\sem{\varphi}\in\valus$.
Let $\qmu$ (respectively, $\qmu_1^{+}$) consist of all 
quantitative $\mu$-calculus formulas (respectively, all 
quantitative $\mu$-calculus formulas with only the $\pre_1$ operator
and all negations before atomic propositions).
The result of \cite{dAMRS07} shows that for all states $s, t \in S$,
\begin{align} \label{logical-charact-metrics}
  \metr{s \priosim_{1} t} 
  & = \sup_{\varphi \in \qmu_1^+} (\sem{\varphi}(s) - \sem{\varphi}(t)) &
  \metr{s \priobis_{g} t} 
  & = \sup_{\varphi \in \qmu} |\sem{\varphi}(s) - \sem{\varphi}(t)| \eqpun .
\end{align}

\smallskip\noindent{\bf Metrics for the discounted quantitative 
$\mu$-calculus.}
We call $\dmu^\dfactor$ the discounted $\mu$-calculus with all
discount parameters $\le \dfactor$.
We define the discounted metrics via an $\dfactor$-discounted metric
transformer $H_{\priosim}^\dfactor: \metrsp \mapsto \metrsp$, defined 
for all $d \in \metrsp$ and all $s, t \in S$ by:
\begin{align}
H_{\priosim_1}^\dfactor(d)(s,t) 
  \; = \; \propdist(s, t) \imax 
    \dfactor \cdot
    \Sup_{k \in C(d)} \bigl( 
    \pre_1(k)(s) - \pre_1(k)(t) \bigr) \eqpun . 
    \label{eq-disc-priosim}
\end{align}
Again, $H_{\priosim_1}^\dfactor$ is continuous and monotonic in the 
lattice $(\metrsp, \leq)$.
The {\em $\dfactor$-discounted simulation metric} $\metr{\priosim_1}^\dfactor$ 
is the least fixpoint of $H_{\priosim_1}^\dfactor$, and the 
{\em $\dfactor$-discounted 
bisimulation metric\/} $\metr{\priobis_1}^\dfactor$ is the least symmetrical 
fixpoint of $H_{\priosim_1}^\dfactor$.
The following result follows easily by induction on the Picard
iterations used to compute the distances \cite{luca-icalp-disc-03};
for all states $s, t \in S$ and a discount factor $\dfactor \in [0, 1)$,
\begin{align} 
& \metr{s \priosim_1 t}^\dfactor \le \metr{s \priosim_1 t} & 
& \metr{s \priobis_1 t}^\dfactor \le \metr{s \priobis_1 t} \eqpun .
\label{theo-undisc-bound}
\end{align}
Using techniques similar to the undiscounted case, we can prove that
for every game structure $\game$ and discount factor $\dfactor\in[0,1)$, 
the fixpoint $\metr{\priosim_i}^\dfactor$ is a directed metric
and $\metr{\priobis_i}^\dfactor$ is a metric, and that
they are {\em reciprocal}, i.e., $\metr{\priosim_1}^\dfactor = 
\metr{\priomis_2}^\dfactor$ and 
$\metr{\priobis_1}^\dfactor = \metr{\priobis_2}^\dfactor$. 
Given the discounted bisimulation metric coincides for the two
players, we write $\metr{\priobis_g}^\dfactor$ instead of
$\metr{\priobis_1}^\dfactor$ and $\metr{\priobis_2}^\dfactor$.
We now state without proof that the discounted $\mu$-calculus
provides a logical characterization of the discounted metric. 
The proof is based on induction on the structure of formulas,
and closely follows the result for the undiscounted case
\cite{dAMRS07}.
Let $\dmu^\dfactor$ (respectively, $\dmu_1^{\dfactor,+}$) consist of all 
discounted $\mu$-calculus formulas (respectively, all discounted 
$\mu$-calculus formulas with only the $\pre_1$ operator
and all negations before atomic propositions).
It follows that for all game structures $\game$ and states $s, t \in S$,
\begin{align} 
  \metr{s \priosim_1 t}^\dfactor 
  & = \sup_{\varphi \in \dmu_1^{\dfactor,+}} (\sem{\varphi}(s) - \sem{\varphi}(t)) &
  \metr{s \priobis_g t}^\dfactor 
  & = \sup_{\varphi \in \dmu^\dfactor} |\sem{\varphi}(s) - \sem{\varphi}(t)|
\label{theo-logical-charact-metrics}
\eqpun .
\end{align}

\smallskip\noindent{\bf Metric kernels.} 
The kernel of the metric $\metr{\priobis_g}$ ($\metr{\priobis_g}^\dfactor$)
defines an equivalence relation $\priobis_g$ ($\priobis_g^\dfactor$) on the states 
of a game structure: 
$s \priobis_g t$ $(s \priobis_g t)^\dfactor$ iff $\metr{s \priobis_g t} = 0$
($\metr{s \priobis_g t}^\dfactor = 0$); 
the relation $\priobis_g$ is called the {\em game bisimulation\/}
relation \cite{dAMRS07} and the relation $\priobis_g^\dfactor$ is called the
{\em discounted game bisimulation\/} relation.
Similarly, we define the {\em game simulation\/} preorder 
$s \priosim_1 t$ as the kernel of the directed metric $\metr{\priosim_1}$, 
that is, $s \priosim_1 t$ iff $\metr{s \priosim_1 t} = 0$.
The {\em discounted game simulation\/} preorder is defined 
analogously.

%% file: average.tex
\section{Bounds for Average and Discounted Payoff Games}
\label{sec-average}

From (\ref{logical-charact-metrics}) it follows that the game bisimulation 
metric provides a tight bound for the difference in valuations
of quantitative $\mu$-calculus formulas.
In this section, we show that the game bisimulation metric also 
provides a bound for the difference in average and discounted value 
of games. 
This lends further support for the game bisimulation
metric, and its kernel, the game bisimulation relation, being the 
canonical game metrics and relations. 

\paragraph{Discounted payoff games.} 

Let $\straa$ and $\strab$ be strategies of player~1 and player~2 respectively.
Let $\alpha \in [0,1)$ be a discount factor.
The {\em $\alpha$-discounted} payoff $v^\alpha_1(s,\straa,\strab)$ for
player~1 at a state $s$ for a variable $r\in\vars$ and the strategies
$\straa$ and $\strab$ is defined as:
\begin{align}
v^\alpha_1(s,\straa,\strab) = 
(1-\alpha) \cdot \sum_{n=0}^\infty \alpha^n \cdot
  \E_s^{\straa,\strab} \bigl( \int{r}(X_n) \bigr),
\label{player-1-disc-reward}
\end{align}
where $X_n$ is a random variable representing the state of the
game in step $n$.
The discounted payoff for player~2 is
defined as $v^\alpha_2(s,\straa,\strab) = -v^\alpha_1(s,\straa,\strab)$.
Thus, player~1 wins (and player~2 loses) the ``discounted sum'' of the
valuations of $r$ along the path, where the discount factor weighs
future rewards with the discount $\alpha$. 
Given a state $s\in S$, we are interested in finding the maximal
payoff $v_i^\alpha(s)$ that player~$i$ can ensure against all opponent
strategies, when the game starts from state $s \in S$. 
This maximal payoff is given by: 
\[
w^\alpha_i(s) =
  \sup_{\pi_i \in \Pi_i} \; \inf_{\pi_{\jj} \in \Pi_{\jj}}
  \rew_i(s,\pi_\ii,\pi_\jj) \eqpun .
\]
%
These values can be computed as the limit of the sequence of
$\alpha$-discounted, $n$-step rewards, for $n \to \infty$. 
For $i \in \set{1,2}$, we define a sequence of
valuations $w^\alpha_i(0)(s)$, $w^\alpha_i(1)(s)$, $w^\alpha_i(2)(s)$,
\ldots as follows: for all $s \in S$ and $n \geq 0$: 
\begin{align} \label{recur-player-1-disc-reward}
w^\alpha_\ii(n+1)(s) 
= (1 - \alpha) \cdot \int{r}(s) + \alpha \cdot \pre_\ii(w^\alpha_i(n))(s)
\eqpun .
\end{align}
where the initial valuation $w^\alpha_i(0)$ is arbitrary. 
Shapley proved that $w^\alpha_i = \lim_{n\to \infty} w^\alpha_i(n)$
\cite{Shapley53}. 

\paragraph{Average payoff games.}

Let $\straa$ and $\strab$ be strategies of player~1 and player~2 respectively.
The {\em  average} payoff $\rew_1(s,\straa,\strab)$ for player~1 at a state 
$s$ for a variable $r\in\vars$ and the strategies $\straa$ and $\strab$ 
is defined as
\begin{equation} \label{eq-average-stra}
\rew_1(s,\straa,\strab) = 
\lim\inf_{n \to \infty} 
\frac{1}{n} \sum_{k=0}^{n-1} \E_s^{\straa,\strab} \bigl(
  \int{r}(X_k) \bigr),
\end{equation}
where $X_k$ is a random variable representing the $k$-th state of the 
game.
The reward for player~2 is $\rew_2(s,\straa,\strab) = 
-\rew_1(s,\straa,\strab)$. 
A game structure $\game$ with  average payoff is called an average
reward game.
The {\em  average value} of the game $\game$ at $s$ for
player $i\in\set{1,2}$ is defined as 
\begin{align*}
\orew_\ii(s) =
  \sup_{\pi_\ii \in \Pi_\ii} \; \inf_{\pi_\jj \in \Pi_\jj}
  \rew_i(s,\pi_\ii,\pi_\jj) 
\eqpun .
\end{align*}
\noindent
Mertens and Neyman established the determinacy of  average
reward games, and showed that the limit of the discounted value of a 
game as all the discount factors tend to $1$ is the same as the 
average value of the game: for all $s \in S$ and $i \in \set{1,2}$, we
have $\lim_{\alpha \to 1} w_\ii^\alpha(s) = \orew_\ii(s)$ \cite{MN81}. 
It is easy to show that the average value of a game is a valuation. 

\paragraph{Metrics for discounted and average payoffs.}

We show that the game simulation metric $\metr{\priosim_1}$ provides
a bound for discounted and long-run rewards.
The discounted metric $\metr{\priosim_1}^\dfactor$ on the other hand
does not provide such a bound as the following example shows.

\psfrag{s}{\textbf{\textit{$s$}}}
\psfrag{t}{\textbf{\textit{$t$}}}
\psfrag{s'}{\textbf{\textit{$s'$}}}
\psfrag{t'}{\textbf{\textit{$t'$}}}
\psfrag{2}{\textbf{\textit{$2$}}}
\psfrag{5}{\textbf{\textit{$5$}}}
\psfrag{2.1}{\textbf{\textit{$2.1$}}}
\psfrag{8}{\textbf{\textit{$8$}}}

\psfrag{1}{\textbf{\textit{$1$}}}
\psfrag{0}{\textbf{\textit{$0$}}}

\begin{figure}[ht]
\centering
{\includegraphics[scale=0.8]{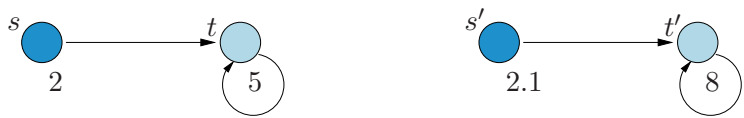}}
\caption{Example that shows that the discounted metric may not be an
upper bound for the difference in the discounted value across states.}
\label{fig:disc-bound}
\end{figure}

\begin{examp}{}
Consider a game consisting of four states $s, t, s', t'$, and a variable
$r$, with $[r](s) = 2$, $[r](s') = 2.1$, $[r](t) = 5$, and $[r](t') =
8$ as shown in Figure~\ref{fig:disc-bound}.
All players have only one move at each state, and the transition
relation is deterministic.
Consider a discount factor $\alpha = 0.9$. 
The $0.9$-discounted metric distance between states 
$s'$ and $s$, is $\metr{s' \priobis_g s}^{0.9} = 0.9 \cdot (8 - 5) = 2.7$.
For the difference in discounted values between the states we proceed
as follows.
Using formulation \ref{recur-player-1-disc-reward}, taking 
$\odrew(0)(t) = 5$, since state $t$ is absorbing, we get
$\odrew(1)(t) = (1 - 0.9) \cdot 5 + 0.9 \cdot 5 = 5$ which leads
to $\odrew(n)(t) = 5$ for all $n \ge 0$.
Similarly $\odrew(n)(t') = 8$ for all $n \ge 0$.
Therefore, the difference in discounted values between $s$ and $s'$, 
again using \ref{recur-player-1-disc-reward}, is given by: 
$\odrew(s') - \odrew(s) = (1 - 0.9) \cdot (2.1 - 2) +
0.9 \cdot (8 - 5) = 2.71$.
\qed
\end{examp}

In the following we consider player~1 rewards (the case for player~2
is identical).

\begin{theo}{} \label{theo-disc-reward-bound}
The following assertions hold.
\begin{enumerate}
\item For all game structures $\game$, $\dfactor$-discounted 
rewards $\odrew_1$, for all states $s, t \in S$, we have,
(a) $\odrew_1(s) - \odrew_1(t) \le [s \priosim_1 t]$ and 
(b) $|\odrew_1(s) - \odrew_1(t)| \le [s \priobis_g t]$.
\item There exists a game structure $\game$, states $s, t \in S$,
such that for all $\alpha$-discounted rewards $\odrew_1$,
$\odrew_1(t) - \odrew_1(s) > \metr{t\priobis_g s}^\dfactor$.
\end{enumerate}
\end{theo}

\begin{proof}
We first prove assertion (1)(a).
As the metric can be computed via Picard iteration, we have for
all $n \geq 0$:
\begin{equation} \label{eq-balls} 
  [s \priosim_1^n t] = \propdist(s, t) \imax 
  \sup_{k \in C([\priosim_1^{n-1}])} (\pre_1(k)(s) - \pre_1(k)(t)) \eqpun .
\end{equation}
We prove by induction on $n \geq 0$ that 
$\odrew_1(n)(s) - \odrew_1(n)(t) \leq [s \priosim_1^n t]$.
For all $s \in S$, taking $\odrew_1(0)(s) = \int{r}(s)$, the base case 
follows. 
Assume the result holds for $n-1 \geq 0$. 
We have: 
\begin{align*}
\odrew_1(n)(s) - \odrew_1(n)(t)
& = (1 - \alpha) \cdot \int{r}(s) + 
    \alpha \cdot \pre_1(w^\alpha(n-1))(s) - \\
&   \quad \ (1 - \alpha) \cdot \int{r}(t) \ - 
    \alpha \cdot \pre_1(w^\alpha(n-1))(t) \\
& = (1 - \alpha) \cdot \bigl(\int{r}(s) - \int{r}(t)\bigr) \ + \\
& \quad \ \ 
    \alpha \cdot \bigl(\pre_1(w^\alpha(n-1))(s) - 
    \pre_1(w^\alpha(n-1))(t)\bigr) \\
&   \le (1 - \alpha) \cdot \propdist(s, t) 
     + \alpha \cdot [s \priosim_1^n t]
    \;\; \le \;\; [s \priosim_1^n t] ,
\end{align*}
where the last step follows by (\ref{eq-balls}), since by the
induction hypothesis we have $\odrew_1(n-1) \in C([\priosim_1^{n-1}])$.
This proves assertion (1)(a).
Given (1)(a), from the definition of $\metr{s \priobis_g t} =
\metr{s \priosim_1 t} \imax \metr{t \priosim_1 s}$,
(1)(b) follows.

The example shown in Figure~\ref{fig:disc-bound} proves the second
assertion.
\end{proof}

Using the fact that the limit of the discounted reward, for a
discount factor that approaches~1, is equal to the average reward, we
obtain that the metrics provide a bound for the difference in average
values as well.  

\begin{cor}{} \label{cor-avg-reward-bound}
For all game structures $\game$ and states $s$ and $t$, we have
(a) $\orew(s) - \orew(t) \leq [s \priosim_1 t]$
and
(b) $|\orew(s) - \orew(t)| \leq [s \priobis_g t]$.
\end{cor}

\paragraph{Metrics for total rewards.}

The {\em total reward} $\trew_1(s,\straa,\strab)$ for player~1
at a state $s$ for a variable $r\in\vars$ and the strategies $\straa
\in \Straa$ and $\strab \in \Strab$ is defined as \cite{FilarVrieze97}: 
\begin{equation} \label{eq-total-stra}
  \trew_1(s,\straa,\strab) = 
  \lim\inf_{n \to \infty} 
  \frac{1}{n} \sum_{k=0}^{n-1} \sum_{j=0}^{k} \E_s^{\straa,\strab} 
  \bigl(\int{r}(X_j) \bigr),
\end{equation}
where $X_j$ is a random variable representing the $j$-th state of the 
game.
The payoff $\trew_2(s,\straa,\strab)$ for player~2 is defined by
replacing  $\int{r}$ with $-\int{r}$ in (\ref{eq-total-stra}). 
The {\em total-reward value} of the game $\game$ at $s$ for
player $i\in\set{1,2}$ is defined analogously to the average
value, via,
\[
\otrew_i(s) =
  \sup_{\pi_i \in \Pi_i} \; \inf_{\pi_{\altro i} \in \Pi_{\altro i}}
  \trew_i(s,\straa,\strab) \eqpun .
\]
While the game simulation metric $[\priobis_g]$ provides an upper
bound for the difference in discounted reward across states,
as well as for the difference in average reward across states, it does
not provide a bound for the difference in total reward. 
We now introduce a new metric, the
{\em total reward metric}, $[\tpriobis_g]$, which provides such a
bound. 
For a discount factor $\dfactor \in [0, 1)$, we define a metric
transformer $H_{\tpriosim_1}^\dfactor: \metrsp \mapsto \metrsp$ as
follows. 
For all $d \in \metrsp$ and $s, t \in S$, we let: 
\begin{align}
& H_{\tpriosim_1}^\dfactor(d)(s,t)
  = \propdist(s, t) + \dfactor \cdot
    \Sup_{k \in C(d)} \bigl( 
    \pre_1(k)(s) - \pre_1(k)(t) \bigr) \eqpun . 
    \label{eq-disc-tpriosim}
\end{align}
The metric $[\tpriosim_1]^\alpha$ (resp.\ $[\tpriobis_1]^\alpha$) is
obtained as the least (resp.\ least symmetrical) fixpoint of
(\ref{eq-disc-tpriosim}). 
We write $[\tpriosim_1]$ for $[\tpriosim_1]^1$, and 
$[\tpriobis_1]$ for $[\tpriobis_1]^1$.
These metrics are {\em reciprocal}, i.e., $\metr{\tpriosim_1}^\dfactor = 
\metr{\tpriomis_2}^\dfactor$ and 
$\metr{\tpriobis_1}^\dfactor = \metr{\tpriobis_2}^\dfactor$. 
If $\dfactor < 1$ we get the {\em discounted total reward metric\/} and
if $\dfactor = 1$ we get the {\em undiscounted total reward metric\/}.
While the discounted total reward metric is bounded, the undiscounted
total reward metric may not be bounded.
The total metrics provide bounds for the difference in discounted, average, 
and total reward between states. 

\begin{theo}{} \label{theo-boundedness-of-total-metrics}
The following assertions hold.
\begin{enumerate}
\item For all game structures $\game$, for all discount factors 
$\dfactor \in [0, 1)$, for all states $s, t \in S$,
\begin{align*}
& (a)\ \metr{s \tpriosim_1 t}^\dfactor \le (\rb - \lb) / (1 - \dfactor), &
& (b)\ \metr{s \tpriosim_1 t}^\dfactor \le \metr{s \tpriosim_1 t}, & \\
& (c)\ \odrew_1(s) - \odrew_1(t) \le \metr{s \tpriosim_1 t}^\dfactor, &
& (d)\ \orew_1(s) - \orew_1(t) \le \metr{s \tpriosim_1 t}, & \\
& (e)\ \otrew_1(s) - \otrew_1(t) \le \metr{s \tpriosim_1 t}.
\end{align*}
\item There exists a game structure $\game$ and states $s, t \in S$
such that,
$\metr{s \tpriosim_1 t} = \infty$.
\end{enumerate}
\end{theo}

\begin{proof}
For assertion (1)(a), notice that $\propdist(s, t) \le (\rb - \lb)$.
Consider the n-step Picard iterate towards the metric distance.
We have,
\[
\metr{s \tpriosim_1^n t}^\dfactor \le
\sum_{i = 0}^{n} \dfactor^i \cdot (\rb - \lb) \eqpun .
\]
In the limit this yields 
$\metr{s \tpriosim_1 t}^\dfactor \le (\rb - \lb) / (1 - \dfactor)$.
Assertion (1)(b) follows by induction on the Picard iterations that
realize the metric distance.
For all $n \ge 0$, 
$\metr{s \tpriosim_1^n t}^\dfactor \le \metr{s \tpriosim_1^n t}$.
Assertion (1)(c) follows by the definition of the discounted total reward 
metric where we have replaced the $\imax$ with a $+$.
By induction, for all $n \ge 0$, 
from the proof of Theorem~\ref{theo-disc-reward-bound} we have,
\[
\odrew_1(n)(s) - \odrew_1(n)(t) \le 
(1 - \dfactor) \cdot \propdist(s, t) + \dfactor \cdot
(\pre_1(\odrew_1(n - 1))(s) - \pre_1(\odrew_1(n-1))(t)) \le
\metr{s \tpriosim_1^n t}^\dfactor \eqpun .
\]
For assertion (1)(d), towards an inductive argument on the Picard iterates 
that realize the metric, for all $n \ge 0$, we have 
$\metr{s \priosim_1^n t} \le \metr{s \tpriosim_1^n t}$,
which in the limit gives 
$\metr{s \priosim_1 t} \le \metr{s \tpriosim_1 t}$.
This leads to 
$\orew_1(s) - \orew_1(t) \le \metr{s \tpriosim_1 t}$,
using Corollary~\ref{cor-avg-reward-bound}.
This proves assertion (1)(d).
We now prove assertion (1)(e) by induction and show that for all 
$n \ge 0$, $\otrew_1(n)(s) - \otrew_1(n)(t) \le \metr{s \tpriosim_1^n t}$.
As the metric can be computed via Picard iteration, we have for all
$n \ge 0$:
\begin{equation} \label{eq-total-balls} 
  [s \tpriosim_1^n t] = \propdist(s, t) + 
  \sup_{k \in C([\tpriosim_1^{n-1}])} (\pre_1(k)(s) - \pre_1(k)(t)) \eqpun .
\end{equation}
We define a valuation transformer 
$u: \valus \mapsto \valus$
as $u(0) = \int{r}$ and for all $n > 0$ and state $s \in S$ as,
\[
u(n)(s) = \int{r}(s) + \pre_1(u(n - 1))(s)
\]
We take $\otrew_1(0) = u(0) = \int{r}$ and for $n > 0$, from the definition of
total rewards (\ref{eq-total-stra}), we get the n-step total reward value 
at a state $s \in S$ in terms of $u$ as,
\begin{align*}
\otrew_1(n)(s) & = 
\frac{1}{n} \cdot \sum_{i = 1}^n u(i)(s)
\eqpun .
\end{align*}
Notice that $\otrew_1(n)(s) \le u(n)$ for all $n \ge 0$.
When $n = 0$, the result is immediate by the definition of $\otrew_1(0)$,
noticing that $\metr{s \tpriosim_1^0 t} = \propdist(s, t)$.
Assume the result holds for $n - 1 \ge 0$.
We have:
\begin{align}
\otrew_1(n)(s) - \otrew_1(n)(t) & = 
\frac{1}{n} \cdot \sum_{i = 1}^n u(i)(s) - 
\frac{1}{n} \cdot \sum_{i = 1}^n u(i)(t) \nonumber \\
& =
\frac{1}{n} \cdot \sum_{i = 1}^n 
(u(i)(s) - u(i)(t)) \nonumber \\
& = 
\frac{1}{n} \cdot \sum_{i = 1}^n
((\int{r}(s) - \int{r}(t)) + \nonumber \\
& \qquad \qquad \;\; (\pre_1(u(i - 1))(s) -
\pre_1(u(i - 1))(t))) \label{pre-total-metric} \\
& \le
\frac{1}{n} \cdot \sum_{i = 1}^n
\metr{s \tpriosim_1^i t} \label{1e-penul} \\
& \le 
\metr{s \tpriosim_1^n t}, \label{1e-ul}
\end{align}
where (\ref{1e-penul}) follows from (\ref{pre-total-metric}) by
(\ref{eq-total-balls}), 
since by our induction hypothesis we have $\otrew_1(i) \le u(i)
\in C(\metr{\tpriosim_1^{i}})$ for all $0 \le i < n$ and
(\ref{1e-ul}) follows from (\ref{1e-penul}) from the monotonicity
of the undiscounted total reward metric.
To prove assertion (2), consider the game structure on the left 
hand side in Figure~\ref{fig:disc-bound}.
The total reward at state $s$ is unbounded; 
$\otrew_1(s) = 2 + 5 + \ldots = \infty$
Now consider a modified version of the game, with identical
structure and with states $s'$ and $t'$ corresponding to
$s$ and $t$ of the original game.
Let $\int{r}(t') = 0$.
In the modified game, $\otrew_1(s') = 2$.
From result (1)(e), since $\otrew_1(s) = \infty$ and
$\otrew_1(s') = 2$, we have $\metr{s \tpriosim_1 s'} = \infty$.
\end{proof}

It is a very simple observation that the quantitative
$\mu$-calculus does not provide a logical characterization for
$[\tpriosim^\alpha_1]$ or $[\tpriosim_1]$. 
In fact, all formulas of the quantitative $\mu$-calculus have
valuations in the interval $[\theta_1, \theta_2]$, while as stated in
Theorem~\ref{theo-boundedness-of-total-metrics}, the total reward can
be unbounded. 
The difference is essentially due to the fact that our version of the
quantitative $\mu$-calculus lacks a ``$+$'' operator. 
It is not clear how to introduce such a $+$ operator in a context
sufficiently restricted to provide a logical characterization for
$[\tpriosim^\alpha_1]$; above all, it is not clear whether a canonical
calculus, with interesting formal properties, would be obtained. 

\subsection{Metric kernels}

We now show that the kernels of all the metrics defined in the paper
coincide: an algorithm developed for the game kernels
$\priosim_1$ and $\priobis_g$, compute the kernels of the
corresponding discounted and total reward metrics as well. 

\begin{theo}{}\label{thm-metric-kernel-canonicity}
For all game structures $\game$, states $s$ and $t$, all discount
factors $\dfactor \in [0, 1)$, the 
following statements are equivalent:
\begin{align*}
& (a) \ \metr{s \priosim_1 t} = 0 & 
& (b) \ \metr{s \priosim_1 t}^\dfactor = 0 & 
& (c) \ \metr{s \tpriosim_1 t}^\dfactor = 0 \eqpun .
\end{align*}
\end{theo}

\begin{proof}
We prove $(a) \Rightarrow (b) \Rightarrow (c) \Rightarrow (a)$.
We assume $0 < \dfactor < 1$.
Assertion $(a)$ implies that $\propdist(s, t) = 0$ and
$\Sup_{k \in C(\metr{\priosim_1})}(\pre_1(k)(s) - \pre_1(k)(t)) \le 0$; 
Since $C(\metr{\priosim_1}^\dfactor) \subs C(\metr{\priosim_1})$ from 
(\ref{theo-undisc-bound}), $(b)$ follows. 
We prove $(b) \Rightarrow (c)$ by induction on the Picard iterations that
compute $\metr{s \priosim_1 t}^\dfactor$ and 
$\metr{s \tpriosim_1 t}^\dfactor$.
The base case is immediate. 
Assume that for all states $s$ and $t$,
$\metr{s \priosim_1^{n - 1} t}^\dfactor = 0$ implies 
$\metr{s \tpriosim_1^{n - 1} t}^\dfactor = 0$.
Towards a contradiction, assume 
$\metr{s \priosim_1^{n} t}^\dfactor = 0$ but
$\metr{s \tpriosim_1^{n} t}^\dfactor > 0$. 
Then there must be $k \in C(\metr{\tpriosim_1^{n - 1}}^\dfactor)$ 
such that $\pre_1(k)(s) - \pre_1(k)(t) > 0$.
By our induction hypothesis, there exists a $\delta > 0$ such that 
$k' = \delta \cdot k \in C(\metr{\priosim_1^{n - 1}}^\dfactor)$.
Since $\pre$ is multi-linear, the player optimal responses in 
$\pre_1(k)(s)$ remain optimal for $k'$.
But this means $(\pre_1(k')(s) - \pre_1(k')(t)) > 0$ for 
$k' \in C(\metr{\priosim_1^{n - 1}}^\dfactor)$, leading to 
$\metr{s \priosim^{n} t}^\dfactor > 0$; a contradiction.
Therefore, $(b) \Rightarrow (c)$.
In a similar fashion we can show that $(c) \Rightarrow (a)$.
\end{proof}

%% file: turnbased.tex
\section{Algorithms for Turn-Based Games and MDPs}
\label{sec-tb-algos}
In this section, we present algorithms for computing the metric
and its kernel for turn-based games and MDPs.
We first present a polynomial time algorithm to compute the operator 
$H_{\priosim_i}(d)$ that gives the {\em exact\/} one-step distance
between two states, for $i \in \set{1,2}$. 
We then present a PSPACE algorithm to decide whether the limit distance
between two states $s$ and $t$ (i.e., $[s \priosim_1 t]$) is at most a 
rational value $r$. 
Our algorithm matches the best known bound known for the special class 
of Markov chains~\cite{vBSW08}.
Finally, we present improved algorithms for the important case of the
kernel of the metrics.
Since by Theorem~\ref{thm-metric-kernel-canonicity} the kernels of the
metrics introduced in this paper coincide, we present our algorithms for
the kernel of the undiscounted metric.
For the bisimulation kernel our algorithm is significantly more
efficient compared to previous algorithms.

\subsection{Algorithms for the metrics}

For turn-based games and MDPs, only one player has a choice
of moves at a given state.
We consider two player~1 states.
A similar analysis applies to player~2 states.
We remark that the distance between states in $S_\ii$ and
$S_\jj$ is always $\rb - \lb$ due to the existence of the 
variable $\turn$.
For a metric $d \in \metrsp$, and states $s, t \in S_1$, computing
$H_{\priosim_1}(d)(s, t)$, given that $p(s, t)$ is trivially computed by 
its definition, entails evaluating the expression,
%
$\Sup_{k \in C(d)} \bigl( \pre_1(k)(s) - \pre_1(k)(t) \bigr)$,
which is the same as,
$\Sup_{k \in C(d)} \Sup_{x \in \dis_1(s)}  \Inf_{y \in \dis_1(t)} 
(\E_s^{x}(k) - \E_t^y(k))$,
%
since $\pre_1(k)(s) = \Sup_{x \in \dis_1(s)} (\E_s^{x}(k))$ and
$\pre_1(k)(t) = \Sup_{y \in \dis_1(t)} (\E_t^{y}(k))$ as player~1
is the only player with a choice of moves at state $s$.
By expanding the expectations, we get the following form,
\begin{multline}\label{eq-tb-one-step}
\Sup_{k \in C(d)}
\Sup_{x \in \dis_1(s)}  \Inf_{y \in \dis_1(t)} 
\biggl(
\sum_{u \in S} \sum_{a \in \mov_1(s)} \trans(s, a)(u) \cdot x(a) \cdot k(u) 
-
\sum_{v \in S} \sum_{b \in \mov_1(t)} \trans(t, b)(v) \cdot y(b) \cdot k(v)
\biggr) \eqpun .
\end{multline}
We observe that the one-step distance as defined in (\ref{eq-tb-one-step})
is a \emph{sup-inf non-linear (quadratic)} optimization problem.
We now present two lemmas by which we transform (\ref{eq-tb-one-step}) to 
an \emph{inf linear} optimization problem, which we solve by linear 
programming (LP).
The first lemma reduces (\ref{eq-tb-one-step}) to an equivalent formulation
that considers only pure moves at state $s$.
The second lemma further reduces (\ref{eq-tb-one-step}), using duality,
to a formulation that can be solved using LP.
\begin{lem}{}\label{lem-tb-a-priori-post-eq}
For all turn-based game structures $\game$, for all player~i states
$s$ and $t$, given a metric $d \in \metrsp$, the following equality holds,
\[
\Sup_{k \in C(d)} \Sup_{x \in \dis_\ii(s)}  \Inf_{y \in \dis_\ii(t)} 
(\E_s^{x}(k) - \E_t^y(k)) =
\Sup_{a \in \mov_\ii(s)}  \Inf_{y \in \dis_\ii(t)} \Sup_{k \in C(d)}
(\E_s^{a}(k) - \E_t^y(k)) \eqpun .
\]
\end{lem}
\proof
We prove the result for player~1 states $s$ and $t$, with the proof
being identical for player~2.
Given a metric $d \in \metrsp$, we have,
\begin{align}
\Sup_{k \in C(d)} \Sup_{x \in \dis_1(s)} \Inf_{y \in \dis_1(t)}
(\E_s^x(k) - \E_t^y(k)) 
& = \Sup_{k \in C(d)} 
(\Sup_{x \in \dis_1(s)} \E_s^x(k) - \Sup_{y \in \dis_1(t)} \E_t^y(k)) 
\nonumber \\
& = \Sup_{k \in C(d)} 
(\Sup_{a \in \mov_1(s)} \E_s^a(k) - \Sup_{y \in \dis_1(t)} \E_t^y(k)) 
\label{pure-pre-decomp} \\
& = \Sup_{k \in C(d)} \Sup_{a \in \mov_1(s)} \Inf_{y \in \dis_1(t)} 
(\E_s^a(k) - \E_t^y(k)) \nonumber \\
& = \Sup_{a \in \mov_1(s)} \Sup_{k \in C(d)} \Inf_{y \in \dis_1(t)} 
(\E_s^a(k) - \E_t^y(k)) \label{pre-swap} \\
& = \Sup_{a \in \mov_1(s)} \Inf_{y \in \dis_1(t)} \Sup_{k \in C(d)}
(\E_s^a(k) - \E_t^y(k)) \label{post-swap}
\end{align}
For a fixed $k \in C(d)$, since pure optimal strategies exist at
each state for turn-based games and MDPs, we replace the 
$\Sup_{x \in \dis_1(s)}$ with
$\Sup_{a \in \mov_1(s)}$ yielding (\ref{pure-pre-decomp}).
Since the difference in expectations is multi-linear, 
$y \in \dis_1(t)$ is a probability distribution 
and $C(d)$ is a compact convex set, we can use the generalized 
minimax theorem \cite{Sio58}, and interchange the innermost 
$\Sup \Inf$ to get 
(\ref{post-swap}) from (\ref{pre-swap}).\qed

The proof of Lemma~\ref{lem-tb-a-priori-post-eq} is illustrated using the
following example.

\psfrag{s}{\textbf{\textit{$s$}}}
\psfrag{t}{\textbf{\textit{$t$}}}
\psfrag{u}{\textbf{\textit{$u$}}}
\psfrag{v}{\textbf{\textit{$v$}}}
\psfrag{a}{\textbf{\textit{$a$}}}
\psfrag{b}{\textbf{\textit{$b$}}}
\psfrag{c}{\textbf{\textit{$c$}}}
\psfrag{e}{\textbf{\textit{$e$}}}
\psfrag{s'}{\textbf{\textit{$s'$}}}
\psfrag{w'}{\textbf{\textit{$w'$}}}
\psfrag{t'}{\textbf{\textit{$t'$}}}
\psfrag{u'}{\textbf{\textit{$u'$}}}
\psfrag{v'}{\textbf{\textit{$v'$}}}
\psfrag{f}{\textbf{\textit{$f$}}}
\psfrag{0}{\textbf{\textit{$0$}}}
\psfrag{1}{\textbf{\textit{$1$}}}

\begin{figure}[ht]
\centering
\subfigure[MDP~1]{
\includegraphics[scale=0.8]{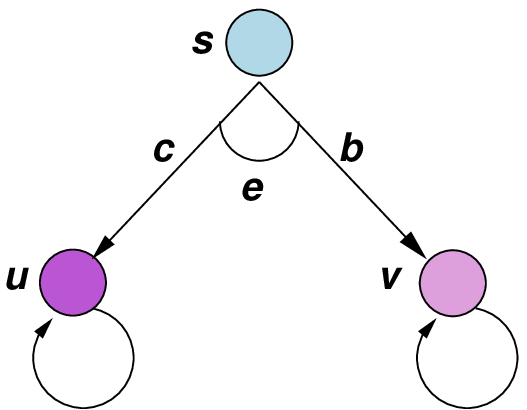}
\label{fig:lem1-mdp1}
} \hspace{1in}
\subfigure[MDP~2]{
\includegraphics[scale=0.8]{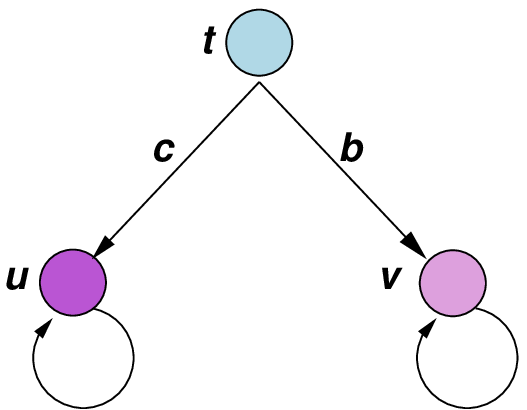}
\label{fig:lem1-mdp2}
}
\caption[]
{An example illustrating the proof of Lemma~\ref{lem-tb-a-priori-post-eq}.}
\label{fig:lem-mdp-metrics}
\end{figure}

\begin{exa}{}
Consider the example in Figure~\ref{fig:lem-mdp-metrics}.
In the MDPs shown in the figure, every move leads to a unique successor state,
with the exception of move $e \in \mov_1(s)$, which leads to
states $u$ and $v$ with equal probability.
Assume the variable valuations are such that all states are at a 
propositional distance of $1$.
Without loss of generality, assume that the valuation $k \in C(d)$
is such that $k(u) > k(v)$.
By the linearity of expectations, for move $c \in \mov_1(s)$, 
$\E_s^c(k) \ge \E_s^x(k)$ for all $x \in \dis_1(s)$.
Similar arguments can be made for $k(u) < k(v)$.
This gives an informal justification for step (\ref{pure-pre-decomp}) in 
the proof; given a $k \in C(d)$, there exist pure optimal strategies for 
the single player with a choice of moves at each state.
While we can use pure moves at states $s$ and $t$ {\em if $k \in C(d)$
is known\/}, the principle difficulty in directly computing the
left hand side of the equality arises from the 
uncountably many values for $k$; the distance is the supremum over all 
possible values of $k$.
In the final equality, step (\ref{post-swap}), and hence by this theorem, 
we have avoided this difficulty, by showing an equivalent expression that 
picks a $k \in C(d)$ to show the difference in distributions induced over 
states.
As we shall see, this enables computing the one-step metric distance
using a trans-shipping formulation.
We remark that while we can use pure moves at state $s$, we cannot do so at 
state $t$ in the right hand side of step (\ref{post-swap}) of the proof.
Firstly, the proof of the theorem depends on $y \in \dis_1(t)$ being convex.
Secondly, if we could restrict our attention to pure moves at state $t$, then 
we can replace $\Inf_{y \in \dis_1(t)}$ with $\Inf_{f \in \mov_1(t)}$ on the 
right hand side.
But this yields too fine a one-step distance.
Consider move $e$ at state $s$.
We see that neither $c$ nor $b$ at state $t$ yield distributions over 
states that match the distribution induced by $e$.
We can then always pick $k \in C(d)$ such that $\E_s^e(k) - \E_t^f(k) > 0$.
If we choose $y \in \dis_1(t)$ such that $y(b) = y(c) = \frac{1}{2}$, we
match the distribution induced by move $e$ from state $s$, which implies 
that for any choice of $k \in C(d)$, 
$\E_s^e(k) - \E_t^{y(b) = y(c) = \frac{1}{2}}(k) = 0$.
Intuitively, the right hand side of the equality can be interpreted
as a game between a protagonist and an antagonist, with the protagonist
picking $y \in \dis_1(t)$, for every pure move $a \in \mov_1(s)$, to
match the induced distributions over states.
The antagonist then picks a $k \in C(d)$ to maximize the difference 
in induced distributions.
If the distributions match, then no choice of $k \in C(d)$ yields a
difference in expectations bounded away from 0.
\end{exa}

From Lemma~\ref{lem-tb-a-priori-post-eq}, given $d \in \metrsp$, we can 
write the player~1 one-step distance between states $s$ and $t$ as follows,
\begin{equation} \label{eq-onestep} 
\OneStep(s,t,d) =
\Sup_{a \in \mov_1(s)}  \Inf_{y \in \dis_1(t)} 
\Sup_{k \in C(d)} (\E_s^{a}(k) - \E_t^y(k)) \eqpun .
\end{equation}
Hence we compute for all $a \in \mov_1(s)$, the expression,
\[
\OneStep(s,t,d,a)=
\Inf_{y \in \dis_1(t)} \Sup_{k \in C(d)} (\E_s^{a}(k) - \E_t^y(k)),
\] 
and then choose the maximum, i.e., 
$\max_{a \in \mov_1(s)} \OneStep(s,t,d,a)$.
We now present a lemma that helps reduce the above
$\Inf-\Sup$ optimization problem to a linear program.
We first introduce some notation.
We denote by $\lambdavec$ the set of variables 
$\lambda_{u,v}$, for $u,v \in S$.
Given $a \in \mov_1(s)$, and a distribution $y \in \dis_1(t)$, we 
write $\lambdavec \in \Phi(s, t, a, y)$ if the 
following linear constraints are satisfied:
\begin{gather*}
\text{(1) for all } v\in S:
\sum_{u \in S} \lambda_{u,v} = \trans(s, a)(v); \quad
\text{(2) for all } u \in S: \sum_{v \in S} 
\lambda_{u,v} = \sum_{b\in \mov_1(t)} y(b)\cdot \trans(t,b)(u); \\
\text{(3) for all } u,v \in S: 
\lambda_{u,v} \ge 0 \eqpun .
\end{gather*}
\begin{lem}{}\label{lem-trans-shipping}
For all turn-based game structures and MDPs $\game$, for all  
$d\in \metrsp$, and for all $s,t \in S$, the following assertion holds:
\[
\sup_{a \in \mov_1(s)} 
\Inf_{y \in \dis_1(t)} \Sup_{k \in C(d)} (\E_s^{a}(k) - \E_t^y(k)) 
=
\sup_{a \in \mov_1(s)} 
\Inf_{y \in \dis_1(t)} \Inf_{\lambdavec \in \Phi(s,t,a,y)} 
\Bigl(
\sum_{u,v \in S} d(u,v) \cdot \lambda_{u,v}
\Bigr) \eqpun . 
\]
\end{lem}
\proof
Since duality always holds in LP, from the LP duality based results of 
\cite{vanBreugelCONCUR01}, for all $a \in \mov_1(s)$ and 
$y \in \dis_1(t)$, the maximization over all $k \in C(d)$ can be 
re-written as a minimization problem as follows:
\[
\Sup_{k \in C(d)} (\E_s^{a}(k) - \E_t^y(k)) 
=
\inf_{\lambdavec \in \Phi(s,t,a,y)} 
\Bigl(
\sum_{u,v \in S} d(u,v) \cdot \lambda_{u,v}
\Bigr) \eqpun .
\]
The formula on the right hand side of the above equality is the 
{\em trans-shipping formulation}, which solves
for the minimum cost of shipping the distribution $\trans(s, a)$ into 
$\trans(t, y)$, with edge costs $d$.
The result of the lemma follows.\qed

\smallskip 

Using the above result we obtain the following LP for $\OneStep(s,t,d,a)$
over the variables:
(a)~$\set{\lambda_{u, v}}_{u, v \in S}$, and 
(b)~$y_b$ for $b \in \mov_1(t)$:
\begin{align} \label{exact-one-step-tb-lp}
\mathrm{Minimize} \quad    \sum_{u, v \in S}d(u,v) \cdot\lambda_{u,v} 
\quad \text{\textrm{subject to}}
\end{align}
\begin{gather*}
\text{(1) for all } v \in S:
\sum_{u \in S} \lambda_{u,v} = \trans(s, a)(v); \qquad
\text{(2) for all } u \in S:
\sum_{v \in S} \lambda_{u,v} = \sum_{b\in \mov_1(t)} y_b\cdot \trans(t,b)(u); \\
\text{(3) for all } u,v\in S: \lambda_{u,v} \ge 0; \qquad
\text{(4) for all } b \in \mov_1(t): y_b \geq 0; \qquad
\text{(5) }  \sum_{b \in \mov_1(t)} y_b=1 \eqpun .
\end{gather*}

\begin{exa}{}
We now use the MDPs in Figure~\ref{fig:mdp1} and \ref{fig:mdp2} to 
compute the simulation distance between states using the results in 
Lemma~\ref{lem-tb-a-priori-post-eq} and
Lemma~\ref{lem-trans-shipping}.
In the figure, states of the same color have a propositional distance
of 0 and states of different colors have a propositional distance
of 1; $\propdist(s, s') = \propdist(t, t') = \propdist(u, u') = 
\propdist(v, v') = \propdist(t', w') = 0$.
In MDP~1, shown in Figure~\ref{fig:mdp1}, $\trans(s, a)(t) =
\trans(t, b)(v) = \trans(t, c)(u) = 1$ and $\trans(t, f)(u) = 
\trans(t, f)(v) = \frac{1}{2}$.
In MDP~2, shown in Figure~\ref{fig:mdp2}, $\trans(s', a)(w') = 
\trans(s', b)(t') = 1$, $\trans(t', c)(u') = \frac{1}{2} - \epsilon$,
$\trans(t', c)(v') = \frac{1}{2} + \epsilon$, 
$\trans(w', e)(u') = \trans(w', f)(v') = 1 - \epsilon$ and
$\trans(w', e)(v') = \trans(w', f)(u') = \epsilon$.

\begin{figure}[ht]
\centering
\subfigure[MDP~1]{
\includegraphics[scale=0.8]{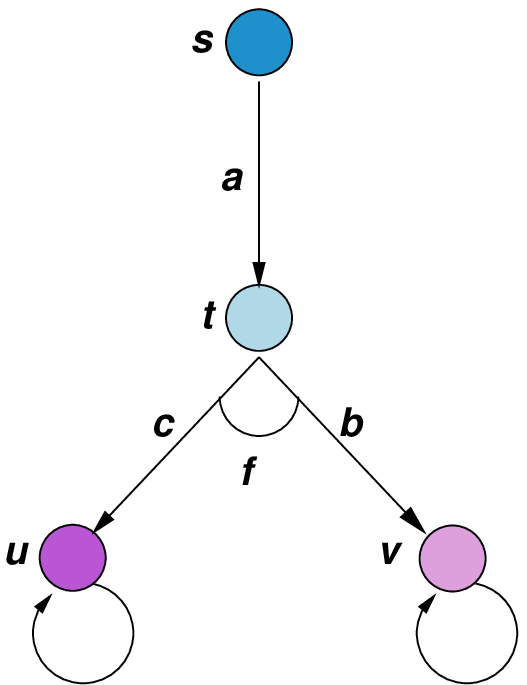}
\label{fig:mdp1}
} \hspace{1in}
\subfigure[MDP~2]{
\includegraphics[scale=0.8]{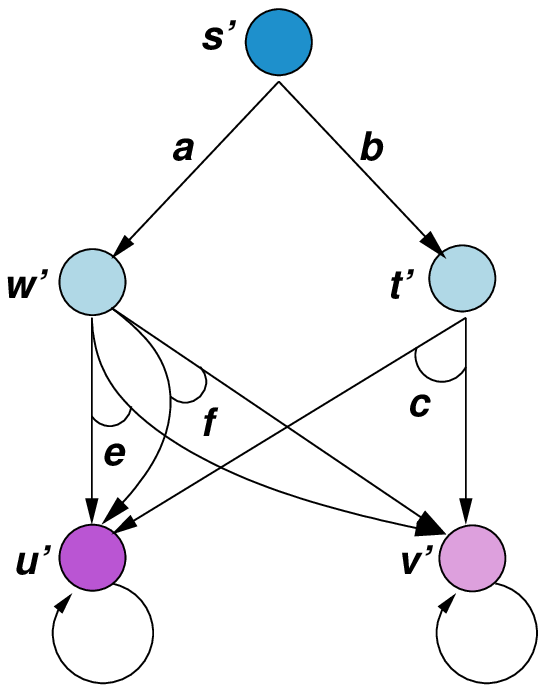}
\label{fig:mdp2}
}
\caption[]
{An example used to compute the simulation metric between states. States
of the same color have a propositional distance of 0.}
\label{fig:mdp-metrics}
\end{figure}

\definecolor{snazz}{rgb}{0.99,0.9,0.9}

\begin{table*}[ht]
\begin{tabular}{|>{\columncolor{snazz}}l|l|l|l|l|}
        \hline
\rowcolor{snazz}
\multicolumn{1}{|>{\columncolor{snazz}}c|}{$t$} & 
\multicolumn{2}{>{\columncolor{snazz}}c|}{$w'$} &
\multicolumn{2}{>{\columncolor{snazz}}c|}{$t'$} \\
\rowcolor{snazz}
$\mov_1(t)\qquad$ & $x \in \dis_1(w')\qquad$ & $Cost\qquad$ 
& $x \in \dis_1(t')\qquad$ & $Cost\qquad$ \\[1.5pt]
        \hline \hline
$b$ & $x(f) = 1$                  & $\epsilon$ 
    & $x(c) = 1$                  & $\frac{1}{2} - \epsilon$ \\[1.5pt]
$c$ & $x(e) = 1$                  & $\epsilon$ 
    & $x(c) = 1$                  & $\frac{1}{2} + \epsilon$ \\[1.5pt]
$f$ & $x(f) = x(e) = \frac{1}{2}$ & $0$ 
    & $x(c) = 1$                  & $\epsilon$ \\[1.5pt]
        \hline
\end{tabular}
\caption{The moves from states $w'$ and $t'$ that minimize the 
trans-shipping cost for each $a \in \mov_1(t)$ and the corresponding
costs.}
\label{table:trans-dist}
\end{table*}
\begin{table*}[ht]
\begin{tabular}{|>{\columncolor{snazz}}l||l|l|l|l|l|}
        \hline
\rowcolor{snazz}
$\metr{\priosim}\qquad$ & $s'\qquad$ & $t'\qquad$ & $w'\qquad$ & $u'\qquad$ & $v'\qquad$ \\[1.5pt]
        \hline \hline
$s$ & $\epsilon$ & $1$                      & $1$        & $1$ & $1$ \\[1.5pt]
$t$ & $1$        & $\frac{1}{2} + \epsilon$ & $\epsilon$ & $1$ & $1$ \\[1.5pt]
$u$ & $1$        & $1$                      & $1$        & $0$ & $1$ \\[1.5pt]
$v$ & $1$        & $1$                      & $1$        & $1$ & $0$ \\[1.5pt]
        \hline
\end{tabular}
\caption{The simulation metric distance between states in MDP~1 and
states in MDP~2.}
\label{table:simdist}
\end{table*}

\begin{figure}[ht]
\centering
\subfigure[$\metr{t \preceq t'} = \frac{1}{2} + \epsilon$]{
\psfrag{v1}{\textbf{\textit{$0$}}}
\psfrag{v0}{\textbf{\textit{$1$}}}
\psfrag{h2}{\textbf{\textit{$\frac{1}{2} + \epsilon$}}}
\psfrag{h1}{\textbf{\textit{$\frac{1}{2} - \epsilon$}}}
\includegraphics[scale=0.8]{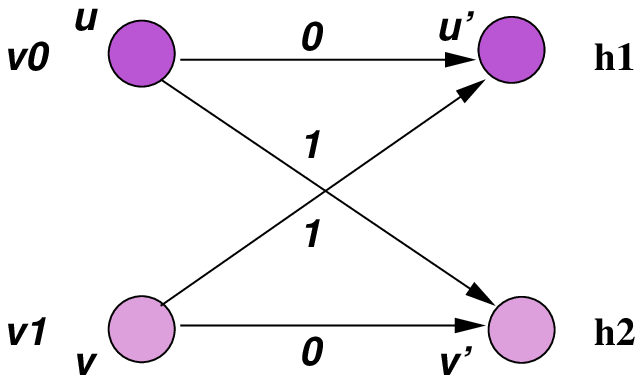}
\label{fig:tt-dist}
} \hspace{1in}
\subfigure[$\metr{s \preceq s'} = \epsilon$]{
\psfrag{v0}{\textbf{\textit{$1$}}}
\psfrag{l1}{\textbf{\textit{$\frac{1}{2} + \epsilon$}}}
\psfrag{l2}{\textbf{\textit{$\epsilon$}}}
\psfrag{h1}{\textbf{\textit{$0$}}}
\psfrag{h2}{\textbf{\textit{$1$}}}
\includegraphics[scale=0.8]{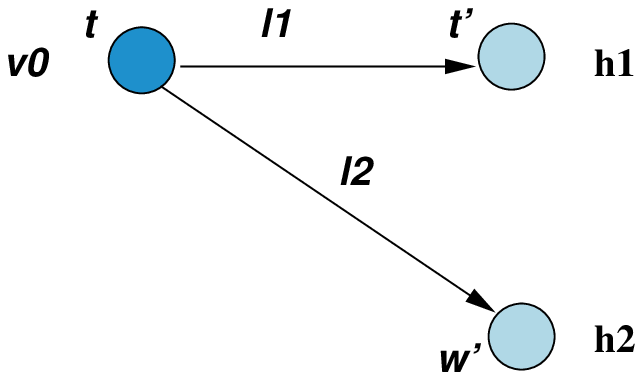}
\label{fig:ss-dist}
}
\label{fig:trans-shipping}
\caption[]
{The trans-shipping formulation that gives the metric distances 
between states.}
\end{figure}

In Table~\ref{table:simdist}, we show the simulation metric distance 
between states of the MDPs in Figure~\ref{fig:mdp1} and 
Figure~\ref{fig:mdp2}.
Consider states $t$ and $t'$.
$c$ is the only move available to player~1 from
state $t'$ and it induces a transition probability of
$\frac{1}{2} + \epsilon$ to state $v'$ and
$\frac{1}{2} - \epsilon$ to state $u'$.
For the pure move $c$ at state $t$, the induced transition 
probabilities and edge costs in the trans-shipping formulation are 
shown in Figure~\ref{fig:tt-dist}.
It is easy to see that the trans-shipping cost in this case is 
$\frac{1}{2} + \epsilon$; shown in Table~\ref{table:trans-dist}
along the row corresponding to move $c$ from state $t$ and 
column corresponding to state $t'$.
Similarly, the trans-shipping cost for the moves $b$ and $f$ from
state $t$ are $\frac{1}{2} - \epsilon$ and $\epsilon$ respectively.
The metric distance $\metr{t \priosim t'}$, which is the maximum 
over these trans-shipping costs is then $\frac{1}{2} + \epsilon$.
Now consider the states $t$ and $w'$.
In Table~\ref{table:trans-dist}, we show for each pure move 
$a \in \mov_1(t)$, the move $x \in \dis_1(w')$ that minimizes the 
trans-shipping cost together with the minimum cost.
In this case it is easy to see that $\metr{t \priosim w'} = \epsilon$.
Given $\metr{t \priosim t'} = \frac{1}{2} + \epsilon$ and 
$\metr{t \priosim w'} = \epsilon$,
we can calculate the distance $\metr{s \priosim s'}$ from the
trans-shipping formulation shown in Figure~\ref{fig:ss-dist}; 
the minimum cost is $\epsilon$ that entails choosing move $a$ from state 
$s'$, giving us $\metr{s \priosim s'} = \epsilon$.
\end{exa}

\begin{thm}{} \label{theo-one-step-poly}
For all turn-based game structures and MDPs $\game$, given $d \in \metrsp$, 
for all states $s, t \in S$, we can
compute $H_{\priosim_1}(d)(s,t)$ in polynomial time by the
Linear Program~(\ref{exact-one-step-tb-lp}). 
\end{thm}

For all states $s, t \in S$, iteration of  $\OneStep(s,t,d)$ converges 
to the exact distance.
However, in general, there are no known bounds for the rate of convergence.
We now present a decision procedure to check whether the 
exact distance between two states is at most a rational value $r$.
We first show how to express 
the predicate $d(s, t) = \OneStep(s,t,d)$.
We observe that since $H_{\priosim_1}$ is non-decreasing, we have
$\OneStep(s,t,d) \geq d(s, t)$. 
It follows that the equality $d(s, t) = \OneStep(s,t,d)$ holds iff 
for every $a \in \mov_1(s)$, of which there are finitely many, all the 
linear inequalities of LP~(\ref{exact-one-step-tb-lp}) are satisfied, 
and $d(s, t)= \sum_{u, v \in S}d(u, v) \cdot\lambda_{u,v}$ holds.
It then follows that $d(s, t) = \OneStep(s,t,d)$ can be written as a 
predicate in the theory of real closed fields. 
Given a rational $r$, two states $s$ and $t$, we present an existential 
theory of reals formula to decide whether $[s \priosim_1 t] \leq r$.
Since $[s \priosim_1 t]$ is the least fixed point of $H_{\priosim_1}$, 
we define a formula $\Phi(r)$ that is true iff, in the fixpoint,
$[s \priosim_1 t] \leq r$,
as follows:
\[
\exists d \in \metrsp. 
[ (\bigwedge_{u, v \in S}\OneStep(u,v,d) = d(u, v)) \land 
(d(s,t) \leq r) ] \eqpun .
\]
If the formula $\Phi(r)$ is true, then there exists a fixpoint $d$,
such that $d(s, t)$ is bounded by $r$, which implies that in the least 
fixpoint $d(s, t)$ is bounded by $r$.
Conversely, if in the least fixpoint $d(s, t)$ is bounded by $r$, then the 
least fixpoint is a witness $d$ for $\Phi(r)$ being true.
Since the existential theory of reals is decidable in PSPACE~\cite{Canny88}, 
we have the following result.

\begin{thm}{(Decision complexity for exact distance).}
For all turn-based game structures and MDPs $\game$, given a rational $r$, 
and two states $s$ and $t$, whether $[s \priosim_1 t] \leq r$ 
can be decided in PSPACE. 
\end{thm}

\smallskip\noindent{\bf{Approximation.}}
Given a rational $\epsilon > 0$, using binary search and 
$\calo(\log(\frac{\rb - \lb}{\epsilon}))$ calls
to check the formula $\Phi(r)$, we can obtain an interval 
$[l,u]$ with $u-l \leq \epsilon$ such that $[s \priosim_1 t]$ 
lies in the interval $[l,u]$.

\begin{cor}{\bf{(Approximation for exact distance).}}
For all turn-based game structures and MDPs $\game$, given a rational 
$\epsilon$, and two states $s$ and $t$, an interval $[l,u]$ with 
$u-l \leq \epsilon$ such that $[s \priosim_1 t] \in [l,u]$ can be 
computed in PSPACE.
\end{cor}

\subsection{Algorithms for the kernel}

The kernel of the simulation metric $\priosim_1$ can be computed as
the limit of the series $\priosim^{0}_1$, $\priosim^{1}_1$,
$\priosim^{2}_1$, \ldots, of relations.
For all $s, t \in S$, we have 
$(s, t) \in \priosim^{0}_1$ iff ${s \loceq t}$.
For all $n \geq 0$, we have $(s, t) \in \priosim^{n+1}_1$ iff 
$\OneStep(s,t,1_{\priosim^n_1}) = 0$. 
Checking the condition $\OneStep(s,t,1_{\priosim^n_1}) = 0$,
corresponds to solving an LP feasibility problem for every 
$a \in \mov_1(s)$, as it suffices to
replace the minimization goal 
$\gamma = \sum_{u, v \in S} 1_{\priosim^n_1}(u, v)\cdot \lambda_{u, v}$
with the constraint $\gamma = 0$ in the LP~(\ref{exact-one-step-tb-lp}).
We note that this is the same LP feasibility problem that was
introduced in \cite{ZhangH07} as part of an algorithm to decide 
simulation of probabilistic systems in which each label may 
lead to one or more distributions over states.

For the bisimulation kernel, we present a more efficient algorithm,
which also improves on the algorithms presented in \cite{ZhangH07}.
The idea is to proceed by partition refinement, as usual for
bisimulation computations. 
The refinement step is as follows: given a partition, two states $s$
and $t$ belong to the same refined partition iff every pure move from
$s$ induces a probability distribution on equivalence classes that can
be matched by mixed moves from $t$, and vice versa. 
Precisely, we compute a sequence $\calq^0$, $\calq^1$, $\calq^2$, 
\ldots, of partitions. 
Two states $s, t$ belong to the same class of $\calq^0$ iff they have
the same variable valuation (i.e., iff ${s \loceq t}$).  
For $n \geq 0$, since by the definition of the bisimulation metric
given in (\ref{eq-game-bis-met}),
$\metr{s \priobis_g t} = 0$ iff $\metr{s \priosim_1 t} = 0$ and
$\metr{t \priosim_1 s} = 0$, two states $s, t$ in a given 
class of $\calq^n$ remain in the same class in 
$\calq^{n+1}$ iff both $(s,t)$ and $(t,s)$ satisfy the set of 
feasibility LP problems $\OneBis(s,t,\calq^n)$ as given below:
\begin{quote}
$\OneBis(s,t,\calq)$ consists of one feasibility LP problem for each
$a \in \mov(s)$.  The problem for $a \in \mov(s)$ has set of variables
$\set{x_b \mid b \in \mov(t)}$, and set of constraints: 
\begin{gather*}
\text{(1) for all } b \in \mov(t): \; x_b \geq 0, \qquad 
\text{(2)~}\sum_{b \in \mov(t)} x_b = 1, \\[1ex]
\text{(3) for all } V \in \calq: \; 
\sum_{b \in \mov(t)} \sum_{u \in V} x_b \cdot \trans(t, b)(u) \geq 
\sum_{u \in V} \trans(s, a)(u) \eqpun .
\end{gather*}
\end{quote}

In the following theorem we show that two states $s, t \in S$ are 
$n + 1$ step bisimilar iff $\OneBis(s, t, \calq^n)$ and
$\OneBis(t, s, \calq^n)$ are feasible.

\begin{thm}{}
For all turn-based game structures and MDPs $\game$, for all
$n \ge 0$, given two states $s, t \in S$ and an $n$-step
bisimulation partition of states $\calq^n$ such that 
$\forall V \in \calq^n$, $\forall u, v \in V$,
$\metr{u \priobis_g v}^n = 0$, the following holds,
\[
\metr{s \priobis_g t}^{n + 1} = 0 \text{ iff }
\OneBis(s, t, \calq^n) \text{ and } \OneBis(t, s, \calq^n) 
\text{ are both feasible}.
\]
\end{thm}

\proof
We proceed by induction on $n$.
Assume the result holds for all iteration steps up to $n$ and
consider the case for $n + 1$.
In one direction, if $\metr{s \priobis_g t}^{n + 1} = 0$, then
$\metr{s \priosim_1 t}^{n + 1} = \metr{t \priosim_1 s}^{n + 1} = 0$
by the definition of the bisimulation metric.
We need to show that given $\metr{s \priosim_1 t}^{n + 1} = 0$,
$\OneBis(s, t, \calq^n)$ is feasible.
The proof is identical for $\metr{t \priosim_1 s}^{n + 1} = 0$.
From the definition of the $n + 1$ step simulation distance,
given $\propdist(s, t) = 0$ by our induction hypothesis, we have,
\begin{align}
\forall b \in \mov_1(s) \Inf_{x \in \dis_1(t)} \Sup_{k \in C(d^n)}
(\E_s^b(k) - \E_t^x(k)) \le 0 
\label{onebis-1}
\eqpun .
\end{align}
Consider a player~1 move $a \in \mov_1(s)$.
Since we can interchange the order of the $\Inf$ and $\Sup$ by the
generalized minimax theorem in 
$\Inf_{x \in \dis_1(t)} \Sup_{k \in C(d^n)}
(\E_s^a(k) - \E_t^x(k))$, the optimal values of $x \in \dis_1(t)$ and 
$k \in C(d^n)$ exist and only depend on $a$.
Let $x_a$ and $k_a$ be the optimal values of $x$ and $k$ that 
realize the $\Inf$ and $\Sup$ in 
$\Inf_{x \in \dis_1(t)} \Sup_{k \in C(d^n)}
(\E_s^a(k) - \E_t^x(k))$.
Using $x_a$ and $k_a$ in (\ref{onebis-1}) we have:
\begin{align}
\E_t^{x_a}(k_a) & \ge \E_s^a(k_a) \nonumber \\
\sum_{u \in S} \trans(t, x_a)(u) \cdot k_a(u) & \ge
\sum_{v \in S} \trans(s, a)(v) \cdot k_a(v) \nonumber \\
\sum_{V \in \calq^n} \sum_{u \in V}
\trans(t, x_a)(u) \cdot k_a(u) & \ge
\sum_{V \in \calq^n} \sum_{v \in V} \trans(s, a)(v) \cdot k_a(v) 
\label{onebis-2} \\
\sum_{V \in \calq^n} \sum_{u \in V} \trans(t, x_a)(u) & \ge
\sum_{V \in \calq^n} \sum_{v \in V} \trans(s, a)(v) \label{onebis-3} \\
\forall V \in \calq^n . \biggl(
\sum_{u \in V} \trans(t, x_a)(u) & \ge
\sum_{u \in V} \trans(s, a)(u) \biggr), \label{onebis-4}
\end{align}
where (\ref{onebis-3}) follows from (\ref{onebis-2}) by noting that
for all $V \in \calq^n$, for all states $u, v \in V$, 
$d^n(u, v) = d^n(v, u) = 0$, by our hypothesis, leading to 
$k(u) - k(v) \le d^n(u, v) = 0$ and 
$k(v) - k(u) \le d^n(v, u) = 0$, which implies $k(u) = k(v)$ for all 
$k \in C(d^n)$.
To show (\ref{onebis-4}) follows from (\ref{onebis-3}), assume towards
a contradiction that there exists a $V' \in \calq^n$ such that 
$\sum_{u \in V'} \trans(t, x_a)(u) < \sum_{u \in V'} \trans(s, a)(u)$.
Then there must be a $V'' \in \calq^n$ such that 
$\sum_{u \in V''} \trans(t, x_a)(u) > \sum_{u \in V''} \trans(s, a)(u)$ 
since $\trans(t, x_a)$ is a probability distribution and the sum
of the probability mass allocated to each equivalence class should be $1$.
Further, for all $V \in \calq^n$, for all $u, v \in V$, we have 
$d^n(u, v) = d^n(v, u) = 0$ and for all $u \in V$ and for all 
$w \in S \setm V$, we have $d^n(u, w) = d^n(w, u) = 1$.
Therefore, we can pick a feasible $k' \in C(d^n)$ such that 
$k'(v) > 0$ for all $v \in V''$ and $k'(v) = 0$ for all other states.
Using $k'$ we get $\E_s^a(k') - \E_t^{x_a}(k') > 0$ which means $k_a$ is 
not optimal, contradicting (\ref{onebis-1}).

In the other direction, assume that $\OneBis(s, t, \calq^n)$ is
feasible.
We need to show that $\metr{s \priosim_1 t}^{n + 1} = 0$.
Since $\OneBis(s, t, \calq^n)$ is feasible, there exists a
distribution $x_a \in \dis_1(t)$ for all $a \in \mov_1(s)$ such
that,
$\forall V \in \calq^n . (\sum_{u \in V} \trans(t, x_a)(u) \ge
\sum_{v \in V} \trans(s, a)(v))$.
By our induction hypothesis, this implies that for all $k \in C(d^n)$, 
we have $(\E_s^a(k) - \E_t^{x_a}(k)) \le 0$ and
in particular $\Sup_{k \in C(d^n)} (\E_s^a(k) - \E_t^{x_a}(k)) \le 0$.
Since $\propdist(s, t) = 0$ by our hypothesis and we have shown,
\[
\forall a \in \mov_1(s) \Inf_{x \in \dis_1(t)} \Sup_{k \in C(d^n)}
(\E_s^a(k) - \E_t^x(k)) \le 0,
\]
we have, from Lemma~\ref{lem-tb-a-priori-post-eq},
\[
\metr{s \priosim_1 t}^{n + 1} = \propdist(s, t) \imax
\Sup_{a \in \mov_1(s)} \Inf_{x \in \dis_1(t)} \Sup_{k \in C(d^n)}
(\E_s^a(k) - \E_t^x(k)) = 0\eqpun .
\] 
In a similar fashion, if $\OneBis(t, s, \calq^n)$ is feasible
then $\metr{t \priosim_1 s}^{n + 1} = 0$, which leads to
$\metr{s \priobis_g t}^{n + 1} = 0$ by the definition of the
bisimulation metric, as required.\qed

\noindent{\bf Complexity.}
The number of partition refinement steps required for the
computation of both the simulation and the bisimulation kernel is
bounded by $O(|S|^2)$ for turn-based games and MDPs, where $S$
is the set of states.
At every refinement step, at most $O(|S|^2)$ state pairs are considered,
and for each state pair $(s,t)$ at most $|\mov(s)|$ 
LP feasibility problems  needs to be solved. 
Let us denote by $\LPF(n,m)$ the complexity of solving the
feasibility of $m$ linear inequalities over $n$ variables.
We obtain the following result.

\begin{thm}{}
For all turn-based game structures and MDPs $\game$, the following assertions 
hold:
\begin{enumerate}[\em(1)]
\item the simulation kernel can be computed in
$\calo\big(n^4 \cdot m \cdot \LPF(n^2 + m,n^2 +2n + m+2)\big) $ time;

\item the bisimulation kernel can be computed in
$\calo\big(n^4 \cdot m \cdot \LPF(m,n + m+1)\big) $ time;

\end{enumerate}
where $n=|S|$ is the size of the state space, and
$m=\max_{s\in S} |\mov(s)|$.
\end{thm}

\begin{rem}{} The best known algorithm for $\LPF(n,m)$ works in time 
$\calo(n^{2.5} \cdot\log(n))$~\cite{Ye06} (assuming each arithmetic
operation takes unit time).
The previous algorithm for the bisimulation kernel checked
two way simulation and hence has the complexity 
$\calo(n^4 \cdot m \cdot (n^2+m)^{2.5} \cdot \log (n^2 +m))$,
whereas our algorithm works in time 
$\calo(n^4 \cdot m \cdot m^{2.5} \cdot \log (m))$.
For most practical purposes, the number of moves at a 
state is constant (i.e., $m$ is constant). 
For the case when $m$ is constant, the previous best known
algorithm worked in $\calo(n^9 \cdot \log (n))$ time, whereas
our algorithm works in time
$\calo(n^4)$.
\end{rem}

%% file: concurrent.tex
\section{Algorithms for Concurrent Games}

In this section we first show that the computation of the metric distance 
is at least as hard as the computation of optimal values in 
concurrent reachability games.
The exact complexity of the latter is open, but it is known to be 
at least as hard as the square-root sum problem, which is in 
PSPACE but whose inclusion in NP is a long-standing open problem 
\cite{EtessamiYannakakis07,GareyGrahamJohnson76}.
Next, we present algorithms based on a decision procedure for
the theory of real closed fields, for both checking the bounds of
the exact distance and the kernel of the metrics.
Our reduction to the theory of real closed fields removes one 
quantifier alternation when compared to the previous known 
formula (inferred from \cite{dAMRS07}).
This improves the complexity of the algorithm.

\input{reduction}

\subsection{Algorithms for the metrics}
We first prove a lemma that helps to obtain
reduced-complexity algorithms for concurrent games. 
The lemma states that the distance
$\metr{s \priosim_1 t}$ is attained by restricting player~2 to pure 
moves at state $t$, for all states $s, t \in S$.

\begin{lem}{}\label{lem-pl2-pure}
For all concurrent game structures $\game$ and all metrics
$d \in \metrsp$, we have,
\begin{multline} \label{eq-a-priori-pure}
\Sup_{k \in C(d)}
  \Sup_{x_1 \in \dis_1(s)} \Inf_{y_1 \in \dis_1(t)}
  \Sup_{y_2 \in \dis_2(t)} \Inf_{x_2 \in \dis_2(s)}
  (\E_{s}^{x_1,x_2}(k)) - \E_{t}^{y_1,y_2}(k)) \\
=
  \Sup_{k \in C(d)}
  \Sup_{x_1 \in \dis_1(s)} \Inf_{y_1 \in \dis_1(t)}
  \Sup_{b \in \mov_2(t)} \Inf_{x_2 \in \dis_2(s)}
  (\E_{s}^{x_1,x_2}(k) - \E_{t}^{y_1,b}(k)) \eqpun .
\end{multline}
\end{lem}
\proof
To prove our claim we fix $k \in C(d)$, and player~1 mixed moves 
$x \in \dis_1(s)$, and $y \in \dis_1(t)$.
We then have,
\begin{align}
\Sup_{y_2 \in \dis_2(t)} \Inf_{x_2 \in \dis_2(s)} 
(\E_{s}^{x,x_2}(k)) - \E_{t}^{y,y_2}(k)) 
& = 
\Inf_{x_2 \in \dis_2(s)} \E_s^{x,x_2}(k) -
\Inf_{y_2 \in \dis_2(t)} \E_t^{y,y_2}(k) \label{inf-decomp} \\
& =
\Inf_{x_2 \in \dis_2(s)} \E_s^{x,x_2}(k) -
\Inf_{b \in \mov_2(t)} \E_t^{y,b}(k) \label{inf-pure-decomp} \\
& =
\Sup_{b \in \mov_2(t)} \Inf_{x_2 \in \dis_2(s)} 
(\E_{s}^{x,x_2}(k) - \E_{t}^{y,b}(k)), \nonumber
\end{align}
where (\ref{inf-pure-decomp}) follows from (\ref{inf-decomp})
since the decomposition on the rhs of (\ref{inf-decomp}) yields two 
independent linear optimization problems; the optimal values are
attained at a vertex of the convex hulls of the distributions induced 
by pure player~2 moves at the two states.
This easily leads to the result.\qed

We now present algorithms for metrics in concurrent games.
Due to the reduction from concurrent reachability games, shown 
in Theorem~\ref{theo-reduction}, it is unlikely that we have 
an algorithm in NP for the metric distance between states.
We therefore construct statements in the theory of real closed fields, 
firstly to decide whether $\metr{s \priosim_1 t} \le r$, for a rational 
$r$, so that we can approximate the metric distance between states $s$ and 
$t$, and secondly to decide if $\metr{s \priosim_1 t} = 0$ in
order to compute the kernel of the game simulation and bisimulation 
metrics.

The statements improve on the complexity that can be achieved by
a direct translation of the statements of \cite{dAMRS07} to the theory
of real closed fields.
The complexity reduction is based on the observation that using
Lemma~\ref{lem-pl2-pure}, we can replace a $\sup$ operator with 
finite conjunction, and therefore reduce the quantifier complexity 
of the resulting formula.
Fix a game structure $\game$ and states $s$ and $t$ of $\game$. 
We proceed to construct a statement in the theory of reals that 
can be used to decide if $\metr{s\priosim_1 t} \le r$, for a given 
rational $r$.

In the following, we use variables $x_1$, $y_1$ and $x_2$ to denote 
a set of variables $\set{x_1(a) \mid a\in\mov_1(s)}$, 
$\set{y_1(a) \mid a\in\mov_1(t)}$ and 
$\set{x_2(b) \mid b\in\mov_2(s)}$ respectively.
We use $k$ to denote the set of variables $\set{k(u)\mid u\in S}$,
and $d$ for the set of variables $\set{d(u,v)\mid u,v\in S}$.
The variables $\alpha, \alpha', \beta, \beta'$ range over reals.
For convenience, we assume $\mov_2(t) = \set{b_1,\ldots, b_l}$.

First, notice that we can write formulas that state that
a variable $x$ is a mixed move for a player at state $s$, and 
$k$ is a constructible predicate (i.e., $k\in C(d)$):
\begin{align*}
\IsDistribution(x,\mov_1(s)) \equiv & 
\bigwedge_{a \in \mov_1(s)} x(a) \ge 0 \wedge
\bigwedge_{a \in \mov_1(s)} x(a) \le 1 \wedge
\sum_{a \in \mov_1(s)} x(a) = 1 \eqpun ,\\
\IsValidValuation(k,d) \equiv & \bigwedge_{u \in S} 
\biggl[
k(u) \ge \theta_1 \wedge k(u) \le \theta_2
\biggr] \wedge
\bigwedge_{u, v \in S} (k(u) - k(v) \le d(u, v)) \eqpun .
\end{align*}
In the following, we write bounded quantifiers of the form
``$\exists x_1\in\dis_1(s)$'' or ``$\forall k \in C(d)$'' which
mean respectively $\exists x_1. \IsDistribution(x_1,\mov_1(s)) \wedge \cdots$
and $\forall k. \IsValidValuation(k,d) \rightarrow \cdots$. 

Let $\eta(k,x_1,x_2,y_1,b)$ be the polynomial
$\E_s^{x_1,x_2}(k) - \E_t^{y_1,b}(k)$.
Notice that $\eta$ is a polynomial of degree $3$.
We write $a = \max\set{a_1,\ldots,a_l}$ for variables $a, a_1,\ldots,a_l$ for
the formula
\[
(a = a_1 \wedge \bigwedge_{i=1}^l a_1\geq a_i) \vee 
\ldots\vee (a = a_l \wedge \bigwedge_{i=1}^l a_l \geq a_i) \eqpun .
\]
We construct the formula for game simulation in stages.
First, we construct a formula $\Phi_1(d, s, t, k, x, \alpha)$ with free 
variables
$d, k, x, \alpha$ such that
$\Phi_1(d,s,t,k,x_1,\alpha)$ holds for a valuation to the variables iff
\begin{align*}
\alpha = \inf_{y_1\in\dis_1(t)}\sup_{b\in\mov_2(t)}\inf_{x_2\in\dis_2(s)}
(\E_s^{x_1,x_2}(k) - \E_t^{y_1,b}(k)) \eqpun .
\end{align*}
We use the following observation to move the innermost $\inf$ ahead of the 
$\sup$ over the finite set $\mov_2(t)$ (for a function $f$):
\[
\sup_{b\in\mov_2(t)}\inf_{x_2\in\dis_2(s)} f(b,x_2,x) = \inf_{x_2^{b_1}\in\dis_2(s)}\ldots\inf_{x_2^{b_l}\in\dis_2(s)}
 \max(f(b_1,x_2^{b_1},x),\ldots,f(b_l,x_2^{b_l},x)) \eqpun .
\]
The formula $\Phi_1(d,s,t,k,x_1,\alpha)$ is given by:
\begin{multline*}
\forall y_1\in\dis_1(t).\forall x_2^{b_1}\in\dis_2(s)\ldots x_2^{b_l}\in\dis_2(s).
\forall w_1\ldots w_l.
\forall a.\forall \alpha'.\\
\exists \hat{y}_1\in\dis_1(t).\exists \hat{x}_2^{b_1}\in\dis_2(s)\ldots \hat{x}_2^{b_l}\in\dis_2(s).
\exists \hat{w}_1\ldots \hat{w}_l.
\exists \hat{a}.
\\
\left[
\begin{array}{ccc}
\left\{\begin{array}{c}
\
\Bigl(w_1=\eta(k,x_1,x_2^{b_1},y_1,b_1)\Bigr)\\
\wedge \cdots \wedge \\
\Bigl(w_l=\eta(k,x_1,x_2^{b_l},y_1,b_l)\Bigr)\wedge\\
\bigl(a = \max\set{w_1,\ldots,w_l}\bigr)
\end{array} 
\right\} &
\rightarrow (a\geq \alpha) 
\end{array}
\right]
\wedge \\
\left[
\begin{array}{ccc}
\left\{\begin{array}{c}
\
\Bigl(\hat{w}_1=\eta(k,x_1,\hat{x}_2^{b_1},\hat{y}_1,b_1)\Bigr)\\
\wedge \cdots \wedge \\
\Bigl(\hat{w}_l=\eta(k,x_1,\hat{x}_2^{b_l},\hat{y}_1,b_l)\Bigr)\wedge\\
\bigl(\hat{a} = \max\set{\hat{w}_1,\ldots,\hat{w}_l} \wedge \hat{a}\geq \alpha'(s,t) \bigr)
\end{array}
\right\} & 
\rightarrow (\alpha \geq \alpha')
\end{array}
\right] \eqpun .
\end{multline*}
Using $\Phi_1$, we construct a formula $\Phi(d, s, t, \alpha)$ with free 
variables 
$d \in \metrsp$ and $\alpha \in \metrsp$ such that
$\Phi(d,s,t,\alpha)$ is true iff:
\[
\alpha = \Sup_{k \in C(d)}
\Sup_{x_1 \in \dis_1(s)} 
\Inf_{y_1 \in \dis_1(t)}
\Sup_{b \in \mov_2(t)} \Inf_{x_2 \in \dis_2(s)}
(\E_{s}^{x_1,x_2}(k) - \E_{t}^{y_1,b}(k)) \eqpun .
\]
The formula $\Phi$ is defined as follows:
\begin{multline} 
\label{phi-for-a-priori}
\forall k\in C(d). \forall x_1\in \dis_1(s).\forall \beta.\forall \alpha'.\\
\biggl[ 
\begin{array}{c}
\Phi_1(d, s, t, k, x_1, \beta) \rightarrow 
  (\beta(s, t) \le \alpha)
\wedge \\
(\forall k'\in C(d). \forall x'_1\in \dis_1(s).\forall\beta'.\Phi_1(d,s,t,k',x'_1,\beta')\wedge \beta'(s, t) \le \alpha')
\rightarrow \alpha \le \alpha'
\end{array}
\biggr] \eqpun .
\end{multline}
Finally, given a rational $r$, we can check if 
$\metr{s \priosim_1 t} \le r$ by checking if the following 
sentence is true:
\begin{align} \label{conc-fixpoint-bound}
\exists d \in \metrsp . \exists a \in \metrsp .
[
(\bigwedge_{u,v \in S} \Phi(d,u,v,a(u, v)) \land (d(u, v) = a(u, v))) \land (d(s,t) \le r)
] \eqpun .
\end{align}
The above sentence is true iff in the least fixpoint, $d(s, t)$ is bounded 
by $r$.
Like in the case of turn-based games and MDPs, given a rational
$\epsilon > 0$, using binary search and 
$\calo(\log(\frac{\rb - \lb}{\epsilon}))$ calls to a decision procedure
to check the sentence (\ref{conc-fixpoint-bound}), we can compute an
interval $[l,u]$ with $u - l \le \epsilon$, such that 
$\metr{s \priosim_1 t} \in [l, u]$.

\smallskip
\noindent{\bf Complexity.}
Note that $\Phi$ is of the form $\forall\exists\forall$, because $\Phi_1$
is of the form $\forall\exists$, and appears in negative position
in $\Phi$.
The formula $\Phi$ has
$(|S|+|\mov_1(s)|+3)$ universally quantified variables, followed by
$(|S|+|\mov_1(s)|+3 + 2(|\mov_1(t)| + |\mov_2(s)|\cdot|\mov_2(t)|+|\mov_2(t)|+2))$
existentially quantified variables, followed by
$2(|\mov_1(t)| + |\mov_2(s)|\cdot |\mov_2(t)| + |\mov_2(t)| +1)$
universal variables.
The sentence (\ref{conc-fixpoint-bound}) introduces $|S|^2 + |S|^2$
existentially quantified variables ahead of $\Phi$.
The matrix of the formula is of length at most quadratic in the size of 
the game, and the maximum degree of any polynomial in the formula 
is $3$.
We define the size of a game $\game$ as:
$|G| = |S| + T$, where
$T = \sum_{s,t \in S} \sum_{a, b \in \moves} |\trans(s,a,b)(t)|$. 
Using the complexity of deciding a formula in the theory of real 
closed fields \cite{Basu99}, which states that a formula with $i$
quantifier blocks, where each block has $l_i$
variables, of $p$ polynomials, has a time complexity bound of
$\calo(p^{\calo(\Pi(l_i + 1))})$, we get the following result.

\begin{thm}{(Decision complexity for exact distance).}
For all concurrent game structures $\game$, given a rational $r$, and two 
states $s$ and $t$, whether $\metr{s \priosim_1 t} \le r$ can be decided in
time $\calo(|G|^{\calo(|G|^5)})$.
\end{thm}
\smallskip\noindent{\bf{Approximation.}}
Given a rational $\epsilon > 0$, using binary search and 
$\calo(\log(\frac{\rb - \lb}{\epsilon}))$ calls
to check the formula {\ref{conc-fixpoint-bound}}, we can obtain an interval 
$[l,u]$ with $u-l \leq \epsilon$ such that $[s \priosim_1 t]$ 
lies in the interval $[l,u]$.

\begin{cor}{\bf{(Approximation for exact distance).}}
For all concurrent game structures $\game$, given a rational $\epsilon$, and 
two states $s$ and $t$, an interval $[l,u]$ with $u-l \leq \epsilon$ 
such that $[s \priosim_1 t] \in [l,u]$ can be computed in time
$\calo(\log(\frac{\rb - \lb}{\epsilon}) \cdot |G|^{\calo(|G|^5)})$.
\end{cor}
In contrast, the formula to check whether $\metr{s \priosim_1 t} \le r$,
for a rational $r$, as implied by the definition of 
$H_{\priosim_1}(d)(s,t)$, that does not use Lemma~\ref{lem-pl2-pure}, has 
five quantifier alternations due to the inner sup, which when combined
with the $2 \cdot |S|^2$ existentially quantified variables in
the sentence (\ref{conc-fixpoint-bound}), yields a decision 
complexity of $\calo(|G|^{\calo(|G|^7)})$.

\subsection{Computing the kernels}
Similar to the case of turn-based games and MDPs, the kernel of the 
simulation metric $\priosim_1$ for concurrent games can be computed as
the limit of the series $\priosim^{0}_1$, $\priosim^{1}_1$,
$\priosim^{2}_1$, \ldots, of relations.
For all $s, t \in S$, we have 
$(s, t) \in \priosim^{0}_1$ iff ${s \loceq t}$.
For all $n \geq 0$, we have $(s, t) \in \priosim^{n+1}_1$ iff 
the following sentence $\Phi_s$ is true:
\begin{align*}
\forall a.\Phi(d^n,s,t,a) \rightarrow a = 0,
\end{align*}
where $\Phi$ is defined as in (\ref{phi-for-a-priori}) and
at step $n$ in the iteration, the distance between any pair of
states $u, v \in S$ is defined as follows,
\begin{align*}
\forall u, v \in S .\  d^n(u, v) = 
\begin{cases}
\; 0 \qquad \text{ if (s, t)} \in \; \priosim^n_1 \\
\; 1 \qquad \text{ if (s, t)} \not\in \; \priosim^n_1
\end{cases} \eqpun .
\end{align*}
To compute the bisimulation kernel, we again proceed by
partition refinement.
For a set of partitions $\calq^0,\calq^1,\ldots$, where
$(s, t) \in V$ for $V \in \calq^n$ implies $(s, t) \in \priobis_1^n$,
$(s,t) \in \priobis^{n+1}$ iff the following sentence $\Phi_b$
is true for the state pairs $(s,t)$ and $(t,s)$:
\begin{align*}
\forall a.\Phi(d^n,s,t,a) \rightarrow a = 0,
\end{align*}
where $\Phi$ is again as defined in (\ref{phi-for-a-priori}) and
at step $n$ in the iteration, the distance between any pair of
states $u, v \in S$ is defined as follows,
\begin{align*}
\forall u, v \in S .\  d^n(u, v) = 
\begin{cases}
\; 0 \qquad \text{ if (s, t)} \in \; \priobis^n_1 \\
\; 1 \qquad \text{ if (s, t)} \not\in \; \priobis^n_1
\end{cases} \eqpun .
\end{align*}

\noindent{\bf Complexity.}
In the worst case we need $\calo(|S|^2)$ partition
refinement steps for computing both the simulation and the
bisimulation relation.
At each partition refinement step the number of state pairs we
consider is bounded by $\calo(|S|^2)$.
We can check if $\Phi_s$ and $\Phi_b$ 
are true using a decision procedure for the theory of real closed fields.
Therefore, we need $\calo(|S|^4)$ decisions to compute the kernels.
The partitioning of states based on the decisions can be done
by any of the partition refinement algorithms, such as 
\cite{PaigeTarjan87}.

\begin{thm}{}
For all concurrent game structures $\game$, states $s$ and $t$,
whether $s \priosim_1 t$ can be decided in $\calo(|G|^{\calo(|G|^3)})$
time,
and whether $s \priobis_g t$ can be decided in $\calo(|G|^{\calo(|G|^3)})$ 
time.
\end{thm}

%% file: reduction.tex
\newcommand{\Prb}{\mathrm{Pr}}

\subsection{Reduction of reachability games to metrics}
We will use the following terms in the result.
A \emph{proposition} is a boolean observation variable, and we say 
a state is labeled by a proposition $q$ iff $q$ is true at $s$.
A state $t$ is \emph{absorbing} in a concurrent game, if both players 
have only one action available at $t$, and the next state of $t$ is 
always $t$ (it is a state with a self-loop).
For a proposition $q$, let $\Diamond q$ denote the set of paths that 
visit a state labeled by $q$ at least once.
In concurrent reachability games, the objective is $\Diamond q$, for a 
proposition $q$, and without loss of generality all states labeled by $q$
are absorbing states.

\begin{theo}{} \label{theo-reduction}
Consider a concurrent game structure $G$, with a single proposition $q$, 
such that all states labeled by $q$ are absorbing states.
We can construct in linear-time a concurrent game structure $G'$, with one 
additional state $t'$, such that for all $s \in S$, we have 
\[
[s \priosim_1 t'] =
\displaystyle
\Sup_{\pi_1 \in \Pi_1} \; \Inf_{\pi_{2} \in \Pi_{2}}\;  \Prb_{s}^{\pi_1,\pi_2}(\Diamond q)
\eqpun .
\]
\end{theo}
\begin{proof}
The concurrent game structure $G'$ is obtained from $G$ by adding an 
absorbing state $t'$.
The states that are not labeled by $q$, and the additional state $t'$,
are labeled by its complement $\neg q$.
Observe there is only one proposition sequence from $t'$, and 
it is $(\neg q)^\omega$.
To prove the desired claim we show that for all $s \in S$ we have
$[s \priosim_1 t'] = \Sup_{\pi_1 \in \Pi_1} \; \Inf_{\pi_{2} \in \Pi_{2}}  
\Prb_s^{\pi_1,\pi_2}(\Diamond q)$.
From a state $s$ in $G$ the possible proposition sequences can be expressed as
the following $\omega$-regular expression: 
$(\neg q)^\omega \cup (\neg q)^* \cdot q^\omega$.
Since the proposition sequence from $t'$ is $(\neg q)^\omega$, the supremum 
of the difference in values over $\qmu$ formulas at $s$ and $t'$ is 
obtained by satisfying the set of paths formalized as $(\neg q)^* \cdot q^\omega$
at $s$.
The set of paths defined as $(\neg q)^* \cdot q^\omega$ is the same as 
reaching $q$ in any number of steps, since all states labeled by $q$ are 
absorbing.
Hence, 
\[
\sup_{\varphi \in \qmu^+} (\sem{\varphi}(s) - \sem{\varphi}(t')) = 
\sem{\mu X.( q \vee \pre_1(X) )}(s)
\eqpun .
\]
It follows from the results of~\cite{dAM04} that for all $s \in S$ we have, 
\[
\sem{\mu X.( q \vee \pre_1(X) )}(s) =
\Sup_{\pi_1 \in \Pi_1} \; \Inf_{\pi_{2} \in \Pi_{2}}  
\Prb_s^{\pi_1,\pi_2}(\Diamond q)
\eqpun .
\]
From the above equalities and the logical characterization result
(\ref{logical-charact-metrics}) we obtain the desired result.
\end{proof}

%% file: discussion.tex
\section{Conclusion: Possible Applications and Open Problems}

We have shown theoretical applications of game metrics with
respect to discounted and long-run average values of games. 
An interesting question regarding game metrics is related to 
their usefulness in real-world applications.
We now discuss possible applications of game metrics.
\begin{itemize}
\item
{\em State space reduction.\/}
The kernels of the metrics are the simulation and bisimulation
relations.
These relations have been well studied in the context of transition
systems with applications in program analysis and verification.
For example, in \cite{KatoenKZJ07} the authors show that bisimulation 
based state space reduction is practical and may result in 
an enormous reduction in model size, speeding up model checking
of probabilistic systems.
\item
{\em Security.\/}
Bisimulation plays a critical role in the formal analysis of
security protocols.
If two instances of a protocol, parameterized by a message $m$,
are bisimilar for messages $m$ and $m'$, then the messages remain
secret \cite{CNP09}.
The authors use bisimulation in probabilistic transition systems to
analyze probabilistic anonymity in security protocols.
\item
{\em Computational Biology.\/}
In the emerging area of computational systems biology, the authors
of \cite{Thorsley09} use the metrics defined in
the context of probabilistic systems 
\cite{DGJP99,vanBreugelCONCUR01,vanBreugel-icalp2001} to compare
reduced models of {\em Stochastic Reaction Networks\/}.
These reaction networks are used to study intra-cellular behavior in 
computational systems biology.
The reduced models are Continuous Time Markov Chains (CTMCs), and
the comparison of different reduced models is via the metric
distance between their initial states.
A central question in the study of intra-cellular behavior is estimating
the sizes of populations of various species that cohabitate cells.
The inter-cellular dynamics in this context is modeled as a 
stochastic process, representing the temporal evolution of the species'
populations, represented by a family $(X(t))_{t \ge 0}$ of random vectors.
For $0 \le i < N$, $N$ being the number of different species,
$X_i(t)$ is the population of species $S_i$ at time $t$.
In \cite{SandmannW08}, the authors show how CTMCs that model system
dynamics can be reduced to Discrete Time Markov Chains (DTMCs) using
a technique called uniformization or discrete-time conversion.
The DTMCs are stochastically identical to the CTMCs and enable 
more efficient estimation of species' populations.
An assumption that is made in these studies is that systems are
spatially homogeneous and thermally equilibrated; the molecules are
well stirred in a fixed volume at a constant temperature. 
These assumptions enable the reduction of these systems to CTMCs and
to DTMCs in some cases.
\end{itemize}
In the applications we have discussed, non-determinism is modeled
probabilistically.
In applications where non-determinism needs to be interpreted
demonically, rather than probabilistically, MDPs or turn-based
games would be the appropriate framework for analysis.
If the interaction between various sources of non-determinism needs
to be modeled simultaneously, then concurrent games would be the
appropriate framework for analysis.
For the analysis of these general models, our results and algorithms
will be useful.

\smallskip\noindent{\bf Open Problems.}
While we have shown polynomial time algorithms for the kernel of the
simulation and bisimulation metrics for MDPs and turn-based games,
the existence of a polynomial time algorithm for the kernel of both
the simulation and bisimulation metrics for concurrent games is an
open problem.
The existence of a polynomial time algorithm to approximate the
exact metric distance in the case of turn-based games and MDPs is 
an open problem.
The existence of a PSPACE algorithm for the decision problem
of the exact metric distance in concurrent games is an open problem.